\documentclass[%
 reprint,
nofootinbib,
 amsmath,amssymb,
 aps,
]{revtex4-2}

\usepackage{graphicx}% Include figure files
\usepackage{dcolumn}% Align table columns on decimal point
\usepackage{bm}% bold math
\usepackage{svg}
\usepackage{soul}
\usepackage{hyperref}% add hypertext capabilities
\hypersetup{
    colorlinks=true,
    linkcolor=blue,
    filecolor=magenta,      
    urlcolor=cyan,
    pdftitle={Overleaf Example},
    pdfpagemode=FullScreen,
    }

\begin{document}

\preprint{APS/123-QED}

\title{Estimating decoding graphs and hypergraphs of memory QEC experiments}

\author{Evangelia Takou}
\email{evangelia.takou@duke.edu}
\affiliation{Duke Quantum Center, Duke University, Durham, NC 27701, USA}
\affiliation{Department of Electrical and Computer Engineering, Duke University, Durham, NC 27708, USA}

\author{Kenneth R. Brown}
\email{ken.brown@duke.edu}
\affiliation{Duke Quantum Center, Duke University, Durham, NC 27701, USA}
\affiliation{Department of Physics, Duke University, Durham, NC 27708, USA}
\affiliation{Department of Electrical and Computer Engineering, Duke University, Durham, NC 27708, USA}
\affiliation{Department of Chemistry, Duke University, Durham, NC 27708, USA}

\date{\today}

\begin{abstract}
Characterizing the error sources of quantum devices is essential for building reliable large-scale quantum architectures and tailoring error correction codes to the noise profile of the devices. Tomography techniques can provide detailed information on the noise or quality of quantum states but are typically both computationally and experimentally intensive with respect to the system's size. For QEC experiments, however, the information captured by a detector error model is sufficient for extracting the success rate of the experiment, as well as some information about the underlying noise. In this work, we estimate Pauli noise on detector error models, including hypergraphs, using only the syndrome statistics. We apply this method to  well-known codes such as the repetition, surface, and 2D color codes. Under bare-ancilla syndrome extraction, two-point correlations are enough to reconstruct the detector error model for repetition or surface codes. For color codes or repetition codes under Steane-style syndrome extraction, we show how to extend the estimation method to multi-point correlators and extract the error rates of the hypergraphs. Finally, we find an increase in logical error suppression when we calibrate the decoder to noise fluctuations typically present in experiments.
\end{abstract}

\maketitle

\section{Introduction}
Scaling up useful quantum computing architectures requires protecting the information against various noise sources using quantum error correction (QEC). 
Several successful QEC experiments have been demonstrated recently, with milestone works achieving below-threshold performance~\cite{GoogleNature2025}, logical teleportation~\cite{ErhardNature2021,Anderson2024arXiv,LacroixArXiv2024}, magic-state distillation~\cite{Rodriguez2024arXiv}, and dynamic error correction~\cite{EickbuschArXiv2024}. 
Decoding the errors of QEC experiments demands fast classical processing and high decoding accuracy to yield higher thresholds. However, there is often a trade-off between these two desirable qualities. At the same time, informing the decoder about potential error configurations or the noise profile can boost its performance. Among the several decoding techniques currently being explored, noise-aware decoding seems promising for improving decoding performance without incurring extra experimental or computational overhead. Noise-aware decoding relies on the knowledge of the devices' noise, which, for example, can be obtained from independent noise characterization experiments such as tomography~\cite{Chuang01111997,CramerNatCommun2010,ArianoPRL2001} or benchmarking~\cite{KnillPRA2008,EisertPRXQ2022,HinesPRX2023}. Additionally, a noise-aware decoder can capture noise drifts~\cite{WangDGR2024arXiv} or exploits device correlations~\cite{ChenPRL2022,WallraffArXiv2025} or noise bias~\cite{FlamiaPRX2019,TuckettPRL2020,ArpitPRXQ2023,Campos2024arXiv} to increase the logical error suppression.

%Need one more paragraph here, maybe specific to work on noise estimation. Cite Flamia's papers, Harper's papers etc

%Add more on challenges and what has been done so far.
To succeed in constructing large-scale quantum computers, noise estimation techniques must also evolve, so that the gap between studying low-level hardware performance and high-level QEC performance is bridged. As the depth and size of the quantum logic circuits increases, so does typically the complexity in characterizing the error mechanisms. From the perspective of QEC, enough information needs to be retained so as not to compromise the decoding performance. State or process tomography techniques can typically provide detailed and rich information about the system's noise, at the cost of becoming cumbersome and practically infeasible for large system sizes~\cite{PanPRL2017}; progress in reducing the cost of these methods, however, has been demonstrated in more efficient variants~\cite{AaronsonArXiv2018,DapengPRL2024,EisertPRL2010,LaflammePRL2017}. In part, these variants can reduce some of the resources for noise estimation but might not be viable in the long term for large-scale QEC experiments. 

Other works, such as ACES~\cite{FlamiaACES2022,HarperPRXQ2025,Harper2025arXiv},  use techniques to learn the Pauli channel eigenvalues and then convert those to the Pauli error rates. This method is generic but  
scales as $4^d$ for a circuit of size $d$. However, for independent noise, Ref.~\cite{HarperPRXQ2025} showed that only $4^2-1$ eigenvalues need to be learned per two-qubit gate, since the only correlations arise from two-qubit gate errors. In the case where ACES is used to learn correlated noise, it also needs to be supplemented with some reasonable graph of interactions~\cite{FlamiaPRXQ2023} such that the noise estimation of the full circuit is partitioned into noise estimation of smaller parts of the circuit, and correlations are truncated to a small region of the device. {\color{black}{One issue that can appear in cycle benchmarking and ACES is that certain Pauli eigenvalues cannot be learned to relative precision~\cite{JiangNatCommun2023, HarperPRXQ2025,Chen2024ArXiv} without assumptions about the noise model~\cite{HarperPRXQ2025,Chen2024ArXiv}, and to sidestep this issue amplification via deeper circuits is needed~\cite{Chen2024ArXiv}, which might lead to runtime issues~\cite{HarperPRXQ2025} in real experiments.   }}

In this paper, we provide an alternative method for learning noise, by focusing on the noise present in decoding graphs or hypergraphs of various QEC codes. In particular, we apply the method of Ref.~\cite{Spitz2018} on decoding graphs, and we further extend it to the case of hypergraphs. By keeping the necessary information required for decoding the logical information, the estimation can be performed in a more scalable manner compared to tomography, provided enough syndrome history is accumulated to reconstruct the statistics. This analysis is not based on any prior information of the noise profile or noise characterization from independent experiments. In particular, we show how to reconstruct the detector error models (DEMs) for repetition, surface, and 2D color codes with bare-syndrome ancilla extraction, where in the latter case we need to estimate a decoding hypergraph. Although Ref.~\cite{WagnerPRR2021} mentioned that the color code does not satisfy the general parameter identifiability for learning independent single-qubit Pauli noise in all error rate ranges, we find that if we restrict our attention to its DEM, then we can learn exactly all the information contained in it. We also estimate the DEM of a repetition code memory under Steane-style syndrome extraction, which is another example of a decoding hypergraph. Additionally, the number of experimental shots we need to estimate a decoding graph or hypergraph does not grow with the system size.  
Lastly, our approach does not require Pauli twirling since we only care about the information that needs to be fed to the decoder. The question of how well the estimated error rates which we use for the decoding match the performance of a decoder where the error rates are obtained by the twirled quantum channel is an open question and an interesting future direction. However, there exist indications in the literature such as Refs.~\cite{GoogleNature2021,GoogleNature2024} where it was observed that simulations with parameters obtained from independent noise characterization reproduce to a good accuracy the edge weights obtained from syndrome statistics of non-twirled circuits, with some discrepancies owed to crosstalk or leakage, or even cosmic ray impacts.

The paper is organized as follows. In Sec.~\ref{Sec:Overview} we give an overview of the estimation problem, discuss previous works and expand the estimation method to hypergraphs. In Sec.~\ref{Sec:Results} we apply our estimation method on various error correction codes with circuit-level noise and compare the decoding performance with the noise-aware model. We further compare the performance of a decoder calibrated to fixed error rates with the performance of a decoder where the error rates vary across qubits and gates. Finally, in Sec.~\ref{Sec:Conclusions} we conclude our results.

\section{Overview of the noise-estimation method for a detector error model \label{Sec:Overview}}

{\color{black}{Optimally decoding errors requires information about the probability of syndromes occurring and their correlation in time obtained from repeated measurements. This is typically achieved by assuming or measuring an error model via independent noise characterization experiments and then calculating the error weights for the decoder graph or hypergraph. For example, Stim~\cite{Gidney2021Stim} takes as an input a previously known 
error model and creates the weighted decoder graph based on it. 
Our approach differs from this logic since we generate the weighted decoder graph from the syndrome measurements alone without relying on prior information. Instead of utilizing experimental resources to obtain a gate or qubit error model first and then translate it into the decoding graph, we directly build the latter.
Here, we will see how to address the problem of learning the error rates of a decoding graph of hypergraph.
}}

%Optimally decoding error syndromes requires information about the probability of obtaining the syndromes and their correlation in time from repeated measurement. This is typically achieved by assuming or measuring an error model and then calculating the error weights on the decoder graph. For example, Stim takes an input error model and generates a weighted decoded graph.  Here we generate the weighted decoded graph from the syndrome measurements alone. This leads to a detector error model instead of a gate or qubit error model.  

A decoding graph is a weighted graph whose nodes are detectors that hold the measurement outcome of an ancilla or the parity addition of multiple outcomes, such that they reconstruct a stabilizer value, or the difference of consecutive measurements of a stabilizer. The edges are error mechanisms that can simultaneously flip the incident nodes. Edges that connect to only one node can further exist, meaning they only flip a single detector and are called boundary edges. The weights of the edges relate to the error probability $p$ of an event via $w = -\ln(p/(1-p))$~\cite{HiggotQuantum2025}. These error probabilities are the quantities that we want to estimate.  Decoding hyper-graphs are defined similarly to decoding graphs, but they also contain hyper-edges. Collectively, we can refer to the decoding graphs and hypergraphs as detector error models~\cite{EisertarXiv2024,Gidney2021Stim,HiggotQuantum2025}.

To learn the error rates of detector error models, we rely solely on the statistics of ancilla measurement outcomes. Those measurement outcomes could be obtained by performing several QEC rounds, $r$, and by repeating the experiment a sufficient number of times, $N$, such that we learn the error rates to a good accuracy. For independent Pauli noise which we are focusing on, each error event follows an independent Bernoulli distribution, and so we expect that the absolute error in the estimation of an error rate scales as $1/\sqrt{N}$. At the same time, it is possible that $N$ needs to also increase as the physical error rates increase to improve the estimation. This is because more configuration degrees of freedom become possible, and so we need a larger number of experimental trials to distinguish them. 

In the case of a decoding graph, all error mechanisms are graph-like, meaning that each error flips at most two detectors. Well-known examples of decoding graphs are the decoding graphs of repetition and surface code memories. Such graphs can be decoded with the minimum-weight perfect matching (MWMP) algorithm~\cite{Edmonds_1965,HiggotACM2022,HiggotQuantum2025}. In terms of the estimation, all graph-like errors can be estimated by correlating at most two detectors that fire simultaneously, whereas, for hypergraphs, this method needs to be extended. We will analyze the estimation procedure we follow in each case shortly.

As we have already mentioned, we will be considering independent Pauli noise in our simulations. However, this does not mean that we cannot have correlations in the noise, either $X,Y,Z$ correlations in a single-qubit depolarizing channel, or correlations between qubits, as in a two-qubit depolarizing channel. For example, the single-qubit depolarizing channel of strength $p$, can be converted to three independent noise channels:
\begin{equation}
    \mathcal{E}_{\text{depol,1},p}(\rho) \mapsto \mathcal{E}_{\sigma_x,p'}\circ\mathcal{E}_{\sigma_y,p'}\circ \mathcal{E}_{\sigma_z,p'}(\rho),
\end{equation}
with the effective probability $p'= 1/2 -\sqrt{1/2-4p/3}$, which is valid for $p\leq 3/4$ (limit of maximal single-qubti depolarization). In this case, each $X$, $Y$, $Z$ error happens independently of the other errors, with an error channel $\mathcal{E}_{\sigma_j,p'}(\rho)=(1-p')\rho + p'\sigma_j\rho\sigma_k$. Similarly, the two-qubit depolarizing channel can be written as:
\begin{equation}
    \mathcal{E}_{\text{depol,2},p}(\rho) \mapsto \prod_{j,k \in \{I,X,Y,Z\}, j,k\neq I,I} \mathcal{E}_{\sigma_j\sigma_k,p'}(\rho),
\end{equation}
with the effective probability $p'=1/2-1/2(1-16p/15)^{1/8}$, which is valid for $p\leq 15/16$ (the maximally mixed limit). We have also defined $\mathcal{E}_{\sigma_j \sigma_k,p'}(\rho)= (1-p')\rho +p' \sigma_j\otimes \sigma_k \rho \sigma_j\otimes \sigma_k$. The conversion of the depolarizing channel as independent composite channels is a mathematical trick which simplifies noise simulations of Pauli channels. In experiments, coherent noise can also exist, which requires a different theoretical treatment and analysis. However, since twirling a quantum channel does not change its average fidelity, the Pauli approximation is valid for simulating stochastic noise~\cite{NIELSENPhysicsLettersA2002}, and further there exist regimes where the effect of coherent noise on the logical error rate is negligible~\cite{DuttonQST2017}. Careful device calibration or dynamical decoupling techniques can also suppress the effects of coherent noise~\cite{LidarSciRep2013,LidarPRL2005,JiangPRA2011}. Other types of correlated noise such as cross-talk (which is typically coherent), can be simulated via a Pauli-twirled incoherent channel, although it is not always an accurate approximation~\cite{zhou2025arXiv}. In our noise estimation simulations we will only consider Pauli channels, and we leave the analysis of other noise sources as a future work. We will further focus on memory QEC experiments, which means our goal is to preserve a logical operator such as the $Z_L$ or $X_L$. The code we developed to produce the results of this paper can be found in Ref.~\cite{Takou2025github}.

\subsection{Two-point correlators \label{Subsec:Two_point}}
We begin by reviewing the method of Ref.~\cite{Spitz2018}, for estimating the error probabilities of a decoding graph, in the case of a repetition code memory. Therein, it was shown that for independent noise, if correlations in the statistics of the syndrome information are truncated up to pairs of detectors, then closed-form solutions describe the error probabilities of the DEM. These formulas are different for boundary or bulk edges, with the error probability of a bulk edge given by:
\begin{equation}\label{Eq:Bulk_Edge}
    p_{ij} = \frac{1}{2}-\sqrt{\frac{1}{4}-\frac{\langle v_iv_j\rangle-\langle v_i \rangle\langle v_j\rangle }{1-2(\langle v_i \rangle +\langle v_j\rangle )+4\langle v_iv_j\rangle}},
\end{equation}
whereas for boundary edges it is given by:
\begin{equation}\label{Eq:Boundary_Edge}
    p_{i}=\frac{1}{2}+\frac{\langle v_i\rangle-1/2}{\prod_{j\neq i}(1-2p_{ij})}.
\end{equation}
The denominator of Eq.~(\ref{Eq:Boundary_Edge}) excludes those edges which are incident to the node $i$. In the above expressions, $\langle v_i\rangle$ is the number of times  detector $i$ fires divided by the number of shots (i.e., experimental trials), and $\langle v_iv_j\rangle$ is the number of times we record coincidences on detectors $i$ and $j$, divided by the number of shots. Once we know the error probabilities, we can then reconstruct the weights of the decoding graph via the equation $w = \text{ln}[(1-p)/p]$. Although the above method was presented for the repetition code, it can be applied to reconstruct any detector error model, provided all errors are graph-like, and under the independent noise model assumption. Further, this method is applicable to code-capacity, phenomenological noise, or circuit-level noise. 

To highlight how this method works, we consider a surface code $X$-memory experiment under phenomenological noise, which means that only data and ancilla qubits experience noise, whereas the gates are assumed to be perfect. 
Although phenomenological noise can be simulated without a circuit-level simulator via Pymatching~\cite{HiggotACM2022,HiggotQuantum2025}, we choose to use Stim~\cite{Gidney2021Stim} to define a circuit where the only errors are input single-qubit depolarizing errors, i.e., single-qubit depolarizing applied on each qubit, at the beginning of every QEC round. Further, for simplicity, we measure only the X-stabilizers and ignore the Z-checks, but our method can also be applied when we consider both types of checks and split the DEM into an X-DEM and a Z-DEM. 

\begin{figure*}
    \centering
    \includegraphics[scale=0.67]{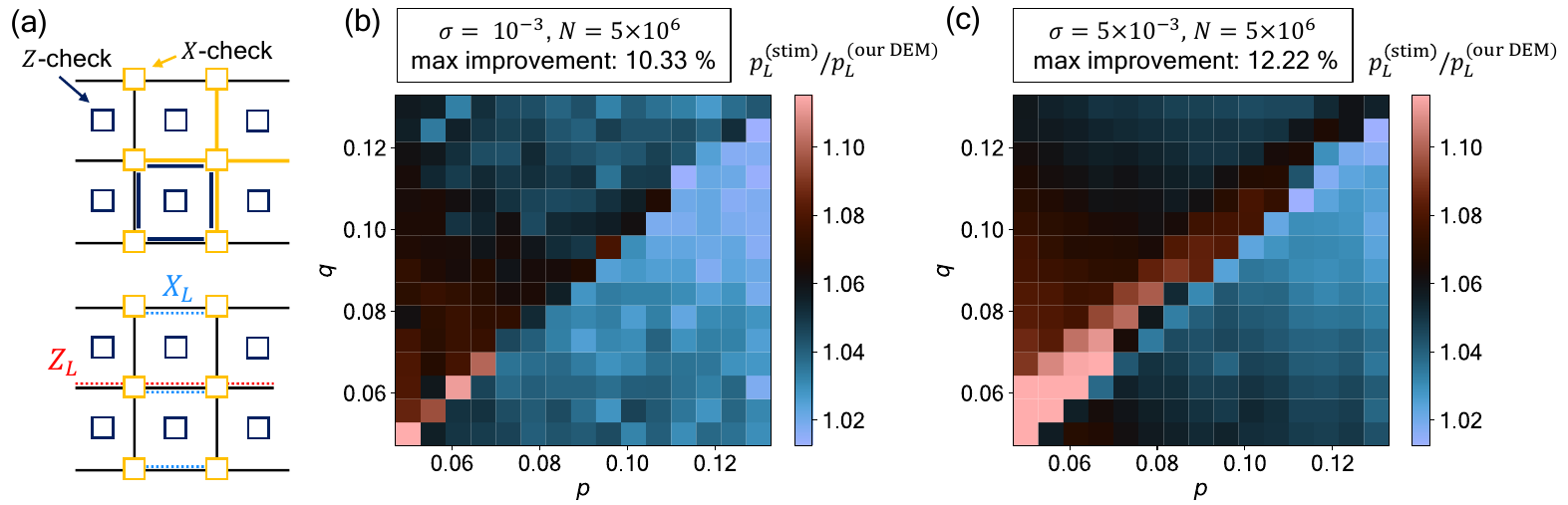}
    \caption{Comparison of logical error rate,  $p_L^{(\text{stim})}$, obtained based on fixed error probabilities $p$ on data qubits and fixed error probabilities $q$ on ancilla qubits, with the logical error rate $p_L^{(\text{our DEM})}$ of a model where the error rates fluctuate across qubits, for a distance $d=3$ surface code memory and $r=3$ QEC rounds. The error rates are sampled from the log-normal distribution with a standard deviation $\sigma$. We consider only the $X$-type stabilizers and decode the $X$-DEM. Our noise model is input single-qubit depolarizing rate on both data and ancilla qubits, and we assume perfect gates. (a) Unrotated surface code lattice of distance $d=3$, with $Z$- and $X$-checks shown as rectangles, forming loop-like and star-like stabilizer operators, respectively. The logical operator $Z_L$ is shown with red, and the logical operator $X_L$ with blue. (b) Ratio of logical error rates as a function of the mean value $p$ of error probabilities of data qubit, and the mean value $q$ of error probabilities of ancilla qubits. The standard deviation for sampling the error probabilities from the log-normal distribution is $\sigma = 10^{-3}$. The maximum improvement we find is $10.33\%$. (c) Same as in (a) for a standard deviation of $\sigma = 5\times 10^{-3}$.  The maximum improvement we find in the logical error rate is $12.22\%$. The number of shots used for the estimation is $N=5\times 10^6$, and the number of shots used to decode each DEM is also $N=5\times 10^6$. }
    \label{fig:Surface_Code_Phenom}
\end{figure*}

In an experimental setup, it is highly likely that the physical error rates vary across the qubits. For this reason, we assume that each qubit experiences single-qubit depolarizing noise that we sample from the log-normal probability distribution with a specified standard deviation. We further assume that the error rates we sample are fixed across the QEC rounds, and across experimental trials. The reason why we sample probabilities from the log-normal probability distribution is because we want them to be non-negative. We consider that the mean $p$ of the log-normal distribution for data qubits  lies in the range $[5\times 10^{-2},0.15]$, and similarly for the mean $q$ of the log-normal distribution for ancilla qubits. To have reproducible results for every value of $p$, $q$ and standard deviation we consider, we set a seed value for generating all the error rates of data qubits, and a different seed value for generating all the error rates of ancilla qubits. We sample 15 evenly spaced intervals for the $p$ and $q$ values, and we further collect the syndrome information for $r=3$ QEC rounds, and $N=8\times 10^5$ shots. We use the collected data to estimate the bulk and boundary probabilities and construct the decoding graph.

The reference model to which we compare is Stim's detector error model, where the error rates are fixed to the mean value of the log-normal distribution with standard deviation set to zero. In other words, for Stim's model we assume that all data qubits experience single-qubit depolarizing of strength $p$, and all ancilla qubits experience single-qubit depolarizing of strength $q$. The actual circuit from which we sample detection events for the decoding is the one were the standard deviation is nonzero, and this is also the circuit that we sample from to learn the actual error rates.

After estimating the error probabilities we decode our estimated model and Stim's model with MWPM, using the Pymatching package~\cite{HiggotQuantum2025,HiggotACM2022}. 
We performed our simulations on an 8-core Apple M1 Chip with 8 GB of RAM. Our results are summarized in Fig.~\ref{fig:Surface_Code_Phenom}, for two different values of standard deviation for the log-normal distribution, i.e, two specific error rate configurations. In Fig.~\ref{fig:Surface_Code_Phenom}(a), we consider a standard deviation of $10^{-3}$ with a mean $p$ for the data qubit error rates, and the same deviation of $10^{-3}$ with a mean $q$ for the ancilla qubit error rates. We also sample $N=5\times 10^6$ shots to collect syndromes for the estimation, and we sample again $N=5\times 10^6$ different shots to decode Stim's DEM and our DEM. The color shows the ratio of Stim's reference logical error rate versus our logical error rate. Across all physical error rates, we find that Stim's logical error rate is larger than the one we estimate. [Due to sampling errors, it could happen, however, that Stim's logical error rate is marginally lower, but the difference in logical error rates can be reduced by increasing the number of shots]. The maximum improvement  factor $|p_L^{(\text{stim})}-p_L^{(\text{our DEM})}|/p_L^{(\text{stim})}\cdot 100\%$ we find across all probability ranges is $10.33\%$. We repeat a similar simulation in Fig.~\ref{fig:Surface_Code_Phenom}(b), but in this case for $\sigma=5\times 10^{-3}$ and $N=5\times 10^6$ shots. We further scale the colorbar ranges to the same minimum and maximum values we found on the left plot, so that we can have a better visual comparison. In this case, we find more regions where Stim's logical error rate is larger than our logical error rate. The maximum improvement obtained for this standard deviation, compared to the reference model, is $12.22\%$. A similar simulation for $d=5$ surface code can be found in Appendix~\ref{App:Surface_Code_Varying_d_5}, where we find a maximum improvement in our logical error rate of $\sim 44\%$.

This phenomenological noise simulation shows that our estimation procedure can take advantage of fluctuations present in experiments, offering a better decoding performance compared to the case of equal error rates across qubits.

\subsection{Extension to higher-order correlations}

So far, we have mentioned that when the statistics are described by up to two-point correlators, we can extract bulk edge probabilities and then use them to redefine the boundary edge probabilities, following a hierarchical approach of subtracting two-point correlations from lower-order events. A similar idea should be applicable when the statistics are described by higher-order correlators (higher than two-point). The two-point estimation method fails to provide the actual two-point contributions because we can no longer separate them from higher-order events. One solution would be to assume that the probability of a higher-order event, for example, an event that flips three detectors, is given by $\langle v_iv_jv_k\rangle$, i.e., the average number of times we see all three detectors click. This would be a good approximation, but only for the case of low error rates, since otherwise, the combination of multiple physical error events can give rise to such a detection pattern, leading typically to underestimation of the actual three-point probability. It is also possible to overestimate the actual three-point probability when several physical errors combine and lead to cancellation effects, i.e., turning off the detectors' state. To make the estimation applicable to any range of physical probabilities, we follow a semi-analytical approach. We form a system of $2^{m}-1$ coupled non-linear equations, given that the DEM contains a single event that flips at most $m$ detectors simultaneously, which we can solve with a numerical optimizer. 

\begin{figure}[!htbp]
    \centering
    \includegraphics[scale=0.9]{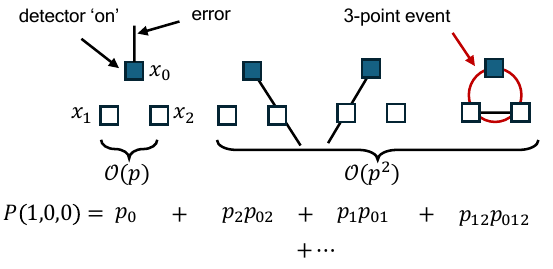}
    \caption{Configuration probability $P(x_0=1,x_1=0,x_2=0)$ of three detectors, written in terms of physical errors. The physical errors include boundary errors, bulk errors, and a three-point event. The equation is truncated to $\mathcal{O}(p^2)$.}
    \label{fig:3_pnt_correlators}
\end{figure}

We highlight here the case of 3-point correlations, which we will further analyze in Sec.~\ref{Subsec:Color_Code}, for the color code memory. The equations that we form correspond to the configuration probabilities $P(x_0,x_1,x_2)$, where each $x_j \in \{0,1\}$. When $x_j=0$, the state of the $j$-th detector is off, whereas when $x_j=1$ the state of the detector is on. For 3-point events, we have $2^3$ equations, namely the combinations ``000'', ``001'', ``010'', ``100'', ``011'', ``101'', ``110'', ``111'' for the outcomes of the detectors. Due to the fact that $\sum_{\{x_j\}}P(\{x_j\})=1$, we can drop one equation due to normalization, for example the equation of $P(0,0,0)$. We write each $P(x_0,x_1,x_2)$ equation in terms of the error probabilities of the DEM that are compatible with the observed state of the detectors. This means that when a detector is on, an odd number of errors flip the detector, whereas an even number of errors (including zero errors) flip any detectors whose state is off.  An example of $P(x_0=1,x_1=0,x_2=0)$ truncated up to $\mathcal{O}(p^2)$ is shown in Fig.~\ref{fig:3_pnt_correlators}, where the red circle denotes the three-point error event.

By performing a memory QEC experiment for some number of shots, we can collect the quantities $\langle v_i\rangle$, $\langle v_iv_j\rangle$, as well as $\langle v_iv_jv_k\rangle$. The last quantity describes the number of coincidences of three detectors divided by the number of shots. Once we have all the expectation values of correlators, we set them equal to the respective configuration probability equation. For example, for the equation $P(x_0=0,x_1=0,x_2=1)$ we need to constraint it to be equal to $\langle v_2\rangle$, whereas for the equation $P(x_0=1,x_1=1,x_2=1)$, we constraint it to be equal to $\langle v_0v_1v_2\rangle$. The equations can be truncated to some order in $p$, depending on the accuracy we want to keep. Typically, for small physical error rates it suffices to keep a low order in $p$, but for higher error rates, we need to increase the truncation order of the equations. Finally, we solve the system of equations with the least-squares method, which gives us the physical error probabilities of the highest-order events, whose contributions we have to subtract from the probabilities of lower-order events, e.g., from boundary or bulk edge probabilities. We will go into more details about this exact procedure in Sec.~\ref{Subsec:Rep_Code_Steane} and Sec.~\ref{Subsec:Color_Code}.

We should further comment that learning a decoding hypergraph is more computationally expensive than learning a decoding graph, since we need to include higher-order correlations in our analysis. However, due to the structure of several QEC codes and the circuit constructions it is typically expected that the order of correlations~\footnote{Note that correlations here refer to detectors that are forced to fire together due to a single error. The type of noise we study throughout the paper corresponds to independent noise. Correlations can further appear due to cross-talk, leakage or even cosmic ray impact.} is a small constant number, given by the maximum number of detectors that are flipped simultaneously in a DEM. A hypergraph DEM needs to be partitioned in such small regions where the numerical equations are solved. The computational cost is mainly owed to the increase in the number of such regions that need to be estimated as the size of the code or the number of QEC rounds increases. However, due to the fact that the estimation can be partitioned in smaller regions allows flexibility in estimating the noise, and potentially targeting regions which might appear to fail more often in experiments.

\section{Application to various codes \label{Sec:Results} under circuit-level noise}
In this section we apply our methodology to estimate the DEM of memory experiments of well-known codes and then decode via MWPM. We also inspect two different syndrome extraction techniques, namely the bare-ancilla syndrome extraction, and the Steane-style syndrome extraction~\cite{SteanePRL1997,Steane2004arXiv,HuangPRL2021,PostlerPRXQuantum2024, HuangSciAdv2024}.

\subsection{Repetition code with bare-ancilla syndrome extraction \label{Subsec:Rep_Code_Bare_Ancilla}}

\begin{figure}[!htbp]
    \centering
    \includegraphics[scale=1]{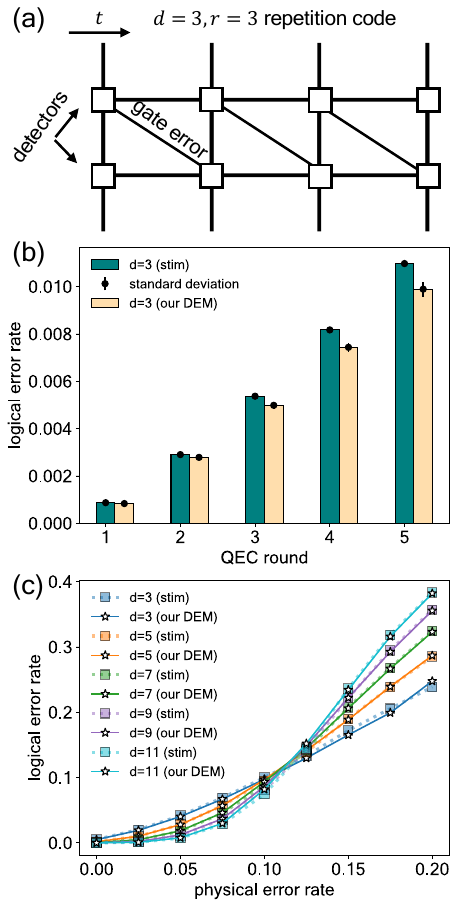}
    \caption{(a) Detector error model for $d=3$ repetition code with bare-ancilla syndrome extraction, and $r=3$ QEC rounds. The diagonal edges represent gate errors, the horizontal edges ancilla errors, and the vertical edges data qubit errors. The last measurement round corresponds to stabilizer reconstruction (measurement of data qubits).
    (b) Logical error rate as a function of the QEC rounds, for $d=3$ repetition code memory, in the limit of only gate errors (two-qubit depolarizing) set to $p_G=0.025$. We use $N=5\times 10^5$ shots for the estimation and decoding, and we repeat the procedure 25 times to extract the standard deviation and average logical error. (c) Logical error rate versus physical error rate for $d$ rounds of a repetition code memory of distance $d$, and under bare-ancilla syndrome extraction. The data and ancilla qubit errors are set to single-qubit depolarizing of equal strength (physical error rate). CNOT gate errors are fixed to two-qubit depolarizing noise of strength $p_G=0.025$. The dashed lines show the performance of Stim's noise-aware model and the solid lines of our reconstructed DEM. We used $10^6$ shots for the estimation and decoding, respectively.  }
    \label{fig:Rep_Code_Bare_Ancilla}
\end{figure}

We begin our analysis with the simplest example of a $Z$-memory repetition code of distance $d$, for which we repeat the syndrome extraction for $r$  QEC rounds. We assume that the syndrome information is obtained via bare-ancilla qubits. The circuit-level detector error model for $d=3,r=3$ repetition code is shown in Fig.~\ref{fig:Rep_Code_Bare_Ancilla}(a), which consist of a 2D grid 
representing data and ancilla qubit errors, as in a phenomenological noise model, whereas the extra diagonal edges represent gate errors. 

To estimate the edge probabilities of the decoding graph, we use the closed-form expressions of Eq.~(\ref{Eq:Bulk_Edge}) and Eq.~(\ref{Eq:Boundary_Edge}). For each boundary edge, we need to include as $p_{ij}$ in the formula of Eq.~(\ref{Eq:Boundary_Edge}) the nearest time edge(s), the nearest space edge, as well as the nearest diagonal edge, if the boundary detector node is incident to a diagonal edge. We define the logical observable as $Z_L=Z_1 \dots Z_d$, which means that any space-bulk, boundary, or diagonal error leads to a flip of the logical observable. On the other hand, time-like errors, i.e., errors on ancilla qubits, do not flip the logical observable. Using all the above information, we reconstruct the DEM and then decode our estimated DEM and Stim's DEM via MWPM.

In Fig.~\ref{fig:Rep_Code_Bare_Ancilla}(b), we show the logical error rate for a $d=3$ repetition code memory as a function of the QEC rounds, assuming we have only a two-qubit depolarizing error after each CNOT gate. We set the error rate to be $p_G=0.025$ and take $N=5\times 10^5$ shots for the estimation and decoding, respectively. We repeat this process 25 times to collect the average logical error rate and the standard deviation. The dark-colored bars show Stim's performance, whereas the light-colored bars show the performance of our DEM. As the number of QEC rounds increases, our reconstructed error model achieves an increasingly improved logical error performance. We observe a similar behavior for $d=5$ and $d=7$ repetition code memory in the limit of only gate errors, although less pronounced than for the case of $d=3$ [see Appendix~\ref{App:Two_qubit_error_limit_rep_code_bare_ancilla}]. 

As a next step, we consider data and ancilla qubit errors in addition to a fixed two-qubit depolarizing gate error, set again to $p_G=0.025$. We assume that data and ancilla qubits experience single-qubit depolarizing error, which, for simplicity, is set to be of the same strength for all qubits. Figure~\ref{fig:Rep_Code_Bare_Ancilla}(c) shows the logical error rate as a function of the physical error rate (single-qubit depolarizing), for various distances $d$, for which we repeat the syndrome extraction for $r=d$ rounds. For each point, we sample for $N=10^6$ shots to perform the estimation and the decoding. The dashed lines show the performance of Stim's noise-aware model, and the solid lines show the performance of our estimated DEM. We notice very good agreement across all physical error rates, even for error rates that exceed the threshold.

\subsection{Surface code with bare-ancilla syndrome extraction \label{Subsec:Surface_Code}}

\begin{figure}[!htbp]
    \centering
    \includegraphics[scale=0.93]{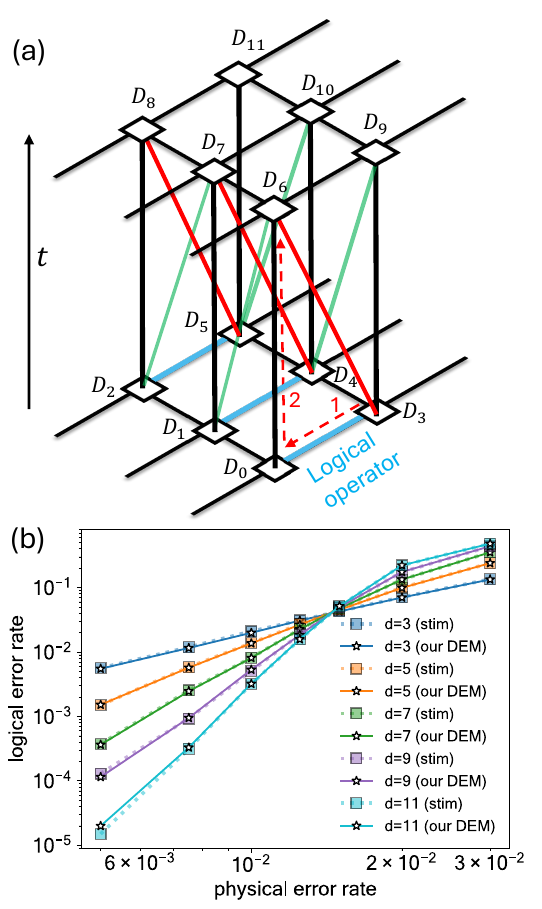}
    \caption{(a) Detector error model of $d=3$ surface code memory. The vertical axis indicates time. The horizontal edges correspond to errors on data qubits, the vertical edges to errors on ancilla qubits, and the diagonal edges correspond to gate errors. The highlighted blue lines correspond to the logical operator. The red diagonal lines flip the logical observable. The path from 1 to 2, shows the decomposition of the diagonal red error as space and time error. (b) Logical error rate as a function of the physical error rate for various distances of $X$-memory surface code. The two-qubit depolarizing rate is equal to the single-qubit depolarizing rate, and the threshold is $1\%$. The dashed lines show the performance of Stim's DEM and the solid lines of our reconstructed DEM. We sampled $200.000$ shots for the estimation and decoding.   }
    \label{fig:Surface_Code}
\end{figure}

We now focus on an $X$-memory surface code, where we measure only the $X$-stabilizers (star-like operators). [A similar analysis holds if we include both $X$- and $Z$-type checks and decode each DEM independently.] Compared to the repetition code memory, we now have a 3D grid with diagonal edges denoting the gate errors, as shown in Fig.~\ref{fig:Surface_Code}. There are two types of diagonal edges, which we show with red and green colors. Each red edge flips the logical observable, $X_L$ (shown as blue edges). The dashed red arrows show one of the two possible paths from the $D_3$ to the $D_6$ detector. In the path shown, we first traverse a space edge (hence, the logical observable is flipped), and then a time edge. 

The error model we assume has input single-qubit depolarizing error on data and ancilla qubits and two-qubit depolarizing error after each CNOT gate. We estimate space-bulk and time edges according to Eq.~(\ref{Eq:Bulk_Edge}). Each boundary edge is estimated via Eq.~(\ref{Eq:Boundary_Edge}) by including those $p_{ij}$ which are incident to the respective detector that checks the boundary edge. For simplicity, we assume that all error rates are equal. We use $N=200.000$ shots to estimate the probabilities and to decode the DEM and compare the performance of our reconstructed DEM to Stim's DEM in Fig.~\ref{fig:Surface_Code}(b). We note very good agreement between the two models, both below and above the threshold of $1\%$. For small physical error rates ($<10^{-2}$) and distances $d\geq 9$, there is slight disagreement in the two logical error rates, which we can resolve this by increasing the number of shots we use for the estimation (for this simulation, we used only 200.000 shots).

\begin{figure}[!htbp]
    \centering
    \includegraphics[scale=0.85]{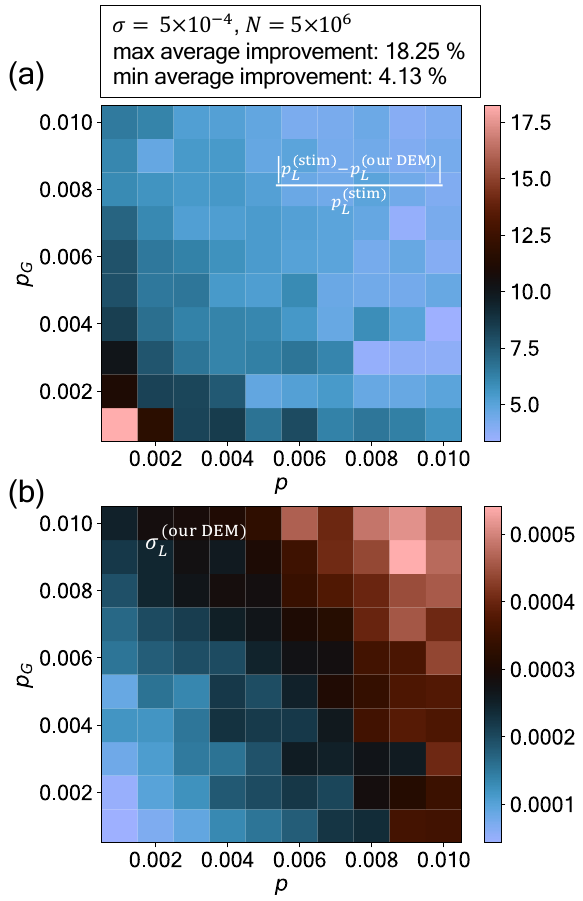}
    \caption{Comparison of logical error rate, $p_L^{(\text{stim})}$, obtained based on fixed single-qubit depolarizing rate $p$ for all qubits, and fixed two-qubit depolarizing rate $p_G$ for gate errors, with the logical error rate, $p_L^{(\text{our DEM})}$, of a model where the error rates for the qubits are sampled from the log-normal distribution with mean $p$, and the gate error rates are sampled from the same distribution with a mean $p_G$. (a) Performance improvement $|p_L^{(\text{stim})}-p_L^{(\text{our DEM})}|/p_L^{(\text{stim})}*100\%$ as a function of the qubit and gate error rates. The standard deviation for the log-normal distributions is set to $\sigma=5\times 10^{-4}$, for both the qubit and gate error rates. The results are averaged over 50 random configurations of error rates per $p$ and $p_G$. The maximum mean improvement is $18.25\%$ and the minimum mean improvement is $4.13\%$. The number of shots used for the estimation and then the decoding is $N=5\times 10^6$. (b) Standard deviation of our logical error rate as a function of $p$ and $p_G$.   }
    \label{fig:Surface_Code_Aver_LER}
\end{figure}

As mentioned in Sec.~\ref{Subsec:Two_point}, the error rates across qubits can vary in experiments, as well as the fidelity of the two-qubit gates across a device. Therefore, it is interesting to test the performance of our method in the presence of varying qubit and gate error rates. For this example, we assume a distance $d=3$ surface code, with $r=3$ QEC rounds. Each data and ancilla qubit experiences input single-qubit depolarizing error that we sample from the log-normal probability distribution with a specified standard deviation, $\sigma$, and the same mean value for ancilla and data qubits, set to $p\in[0.001,0.1]$. Additionally, we sample two-qubit depolarizing gate errors from the log-normal probability distribution with a mean $p_G\in[0.001,0.01]$. We set the standard deviation to $\sigma = 5\times 10^{-4}$ and assume that each error rate remains fixed across QEC rounds. The number of shots we choose to perform the estimation and the decoding is $N=5\times 10^6$. Our goal is to compare the logical error rate we obtain based on our model, where the error rates exhibit fluctuations, compared to a model where the error rates are fixed to the mean (zero standard deviation). To collect the average performance over random configurations, we repeat the calculation 50 times for each $(p,p_G)$ set.

Figure~\ref{fig:Surface_Code_Aver_LER}(a) shows the logical error improvement we obtain over Stim's fixed error rate model as a function of the mean qubit error rate $p$, and the mean gate error $p_G$. The maximum average performance we find is $18.25\%$, and the minimum average performance is $4.13\%$. The best performance occurs for $p,p_G\leq  0.005$, since for such ranges, variations in the error rates with $\sigma=5\times 10^{-4}$ have a bigger impact. Figure~\ref{fig:Surface_Code_Aver_LER}(b) shows the standard deviation of the logical error rate we obtained  by decoding our DEM across the 50 configurations. The maximum standard deviation in the logical error rate is $\sim 5.12\times 10^{-4}$, and the minimum is $\sim 4.27\times 10^{-5}$. We also note that the maximum standard deviation occurs for the bigger values of $p$ and $p_G$, which is expected since for larger physical error rates, we need to increase the number of shots we use for the estimation and the number of shots used for the decoding to reduce this spread. Nevertheless, even in those ranges of bigger error rates, we still obtain better performance under noise fluctuations across qubits and gates compared to a model where the error rates are constant.

\subsection{Repetition code with Steane-style syndrome extraction \label{Subsec:Rep_Code_Steane}}

\begin{figure}[!htbp]
    \centering
    \includegraphics[scale=0.8]{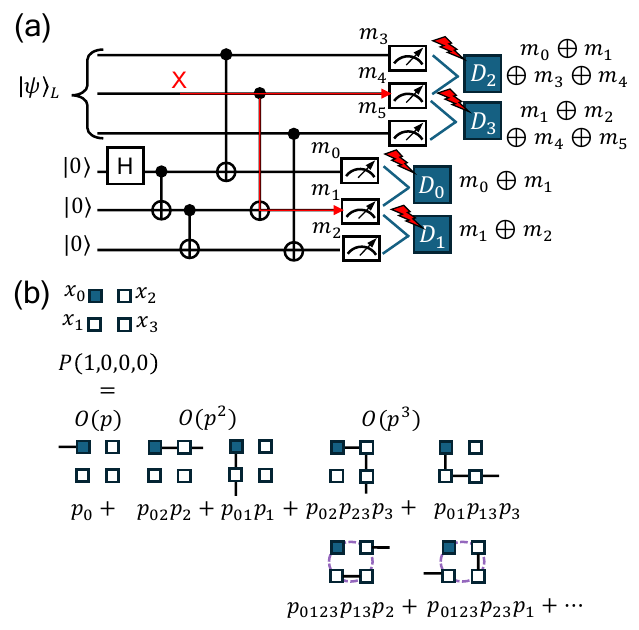}
    \caption{(a) Repetition code circuit of distance $d=3$ and one QEC round,  under Steane-style syndrome extraction. $D_j$ correspond to detectors. One possible error that can flip all four detectors is the $X$ error happening on the second data qubit before the CNOT gate. (b) Terms up to $\mathcal{O}(p^3)$ for the configuration probability equation $P(1,0,0,0)$.}
    \label{fig:Circuit_Steane_SE_Rep_Code}
\end{figure}

Steane error correction exploits the fact that transversal CNOTs performed between two code blocks encoded in the same CSS code lead to a logical CNOT$_L$ between them \cite{SteanePRL1997}. While a logical CNOT$_L$ leaves the logical state $|\psi\rangle_L |+\rangle_L$ invariant, data qubit errors are forced to propagate to the ancilla qubits through individual CNOT gates. This construction satisfies fault-tolerant conditions since one error from the data qubit block propagates as a single error on the ancilla qubit block. Under Steane-style syndrome extraction, the repetition code syndrome is extracted using an ancilla encoded into a GHZ state~\cite{PostlerPRXQuantum2024}, which typically needs to be verified via flag qubits to guarantee fault-tolerance. Here, we consider the $Z$-memory repetition code of distance $d$, with $d$ rounds of syndrome extraction. Detectors are formed analogously to the bare-ancilla syndrome extraction, with the difference that now we need to store in a particular detector $i\in[0,d-2]$ the value $m_i^{(r)}\oplus m_{i+1}^{(r)}\oplus m_{i}^{(r-1)}\oplus m_{i+1}^{(r-1)}$ (prior to the 1st QEC round, it holds that $m_i^{(0)}=0, \forall i$). Due to this definition of detectors, an error that propagates on any of the intermediate ancillas (excluding the first or last) can flip at most four detectors. Thus, we run into the issue that we cannot distinguish a 4-point event from two disjoint 2-point events that happen simultaneously (either time-like or space-like errors). An example of a physical error that flips four detectors is shown in Fig.~\ref{fig:Circuit_Steane_SE_Rep_Code}~(a). 

To address this noise-assignment problem, we include four-point correlators in our estimation method. We form a system of $4^2-1$ equations by expressing the configuration probabilities $P(x_0,x_1,x_2,x_3)$, where $x_j \in\{0,1\}, \forall j$, in terms of combinations of physical error events that are compatible with the observed state of detectors. Each $x_j$ corresponds to the on/off state that each detector can take, with the value of 1 happening when an odd number of errors trigger the particular detector. All detector outcome configurations on four points are $\{0,1\}^4$, and we can drop the $P(0,0,0,0)$ equation using normalization, since $\sum_{\textbf{x}}P(\textbf{x})=1$. An example of $P(1,0,0,0)$ where we keep terms up to $\mathcal{O}(p^3)$, is shown in Fig.~\ref{fig:Circuit_Steane_SE_Rep_Code}~(b). In this particular detector error model, and assuming four detectors $D_i$, $D_{i+1}$, $D_{j}$, $D_{j+1}$, (where $i$ and $j$ refer to detectors differing by one round) we can have the following error events:
\begin{itemize}
    \item Boundary errors: $p_{i}, p_{i+1}, p_{j}, p_{j+1}$
    \item Space-bulk errors: $p_{i,i+1}, p_{j,j+1}$
    \item Time-like errors: $p_{i,j}, p_{i+1,j+1}$
    \item Four-point event: $p_{i,i+1,j,j+1}$.
\end{itemize}

The numerical system of 15 equations needs to be solved for each set of four detectors, enclosing a four-point event. We solve these equations via least-squares optimization and obtain the probabilities of the various error mechanisms. The probabilities we extract as solutions for two-point or boundary errors are not yet correct due to the four-point contributions they include. However, the four-point error probability we find is accurate and can be used to refine all other bulk or boundary edge probabilities. We have two options for the initial values we start with for the bulk and boundary probabilities; we can either use the numerical solutions we found, or we can estimate them via the two-point method of Sec.~\ref{Subsec:Two_point}. Here, we choose the former approach, but either one would be valid. Finally, to remove the contribution of four-point events from bulk and boundary probabilities, we redefine their initial values via the equation:
\begin{equation}
    p_{\text{new}}=(p_{\text{old}}-p_{ijkl})/(1-2p_{ijkl}).
\end{equation}
This equation is recursively applied until we have exhausted all four-point events where the detectors of a single- or two-point event participate. To understand why the above equation holds, we need to remember that we assume an independent noise model. For instance, if we have an event happening with probability $p$ and another independent event happening with probability $q$, which both flip the same set of detectors, then the probability of the detectors firing is $\alpha=p(1-q)+q(1-p)$. In our case, $q$ is the four-point probability and $p$ is what we need to solve for, given that we have estimated $\alpha$ with the two-point estimation method. To give an example, if we want to calculate the probability of the error event that flips $D_0$ and $D_1$, then there is only one four-point event we need to exclude, namely the event $D_0-D_1-D_{0+d-1}-D_{1+d-1}$.

For small error rates, we can truncate the equations to low order in $p$ and estimate the error probabilities with high accuracy. As the error rates increase, we need to increase the truncation order. In the results we present below, we truncate the numerical system that we solve to $\mathcal{O}(p^7)$, which gives us very good agreement across a very broad range of probabilities.

Note that from the perspective of our estimation procedure, we can exclude the flag qubits (for any $d$) and perform the estimation without any post-selection. After we decode, however, we expect to find a lower threshold compared to the case where we verify the ancilla state and post-select on the outcome of all flag qubits being 0. For this present analysis we do not include flag qubits, but our methods are also applicable when flag qubits are incorporated.

\begin{figure}[!htbp]
    \centering
    \includegraphics[scale=0.9]{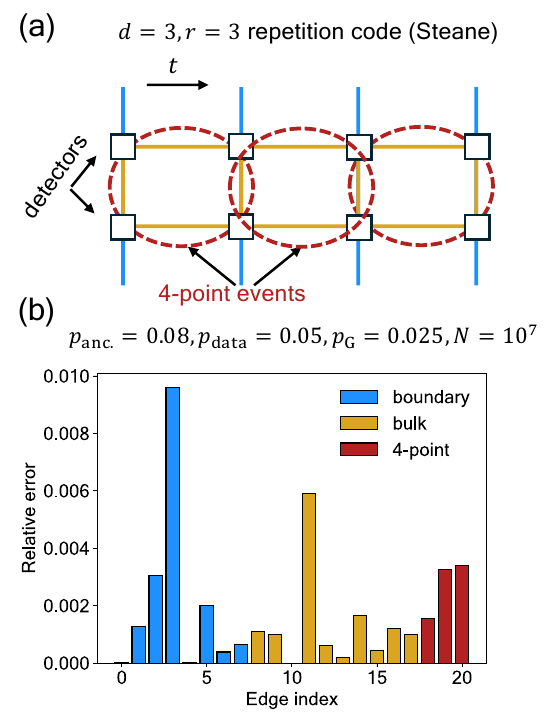}
    \caption{(a) Detector error model for $d=3$ repetition code memory under Steane-style syndrome extraction and for $r=3$ QEC rounds. The last round is the stabilizer reconstruction (measurement of data qubits). The red circles correspond to errors that flip the four detectors they cross. (b) Relative error for each edge, obtained by solving the numerical equations. The blue bars correspond to boundary-edge errors, the yellow bars to bulk errors,  and the red bars to the four-point errors.    }
    \label{fig:Rel_Edge_Error_Steane_SE_Rep_Code}
\end{figure}

To demonstrate the accuracy to which we obtain the error probabilities of the DEM, we begin with a distance $d=3$ repetition code and $r=3$ QEC rounds. We set the single-qubit depolarizing error rate on ancilla qubits to be $p_{\text{anc}.}=0.08$, and on data qubits to be $p_{\text{data}}=0.05$. We additionally consider two-qubit depolarizing noise after each CNOT of strength $p_{\text{G}}=0.025$. We sample the circuit $10^7$ times to collect statistics and then solve for all the error probabilities as we described above. In Fig.~\ref{fig:Rel_Edge_Error_Steane_SE_Rep_Code}(b), we show the relative error rate in estimating each edge of the DEM. With red bars, we show the error in estimating the probabilities of the four-point events, which is below $0.4\%$. We also show the relative error for bulk (space and time) or boundary edges, which is also $<1\%$.

\begin{figure}[!htbp]
    \centering
    \includegraphics[scale=0.78]{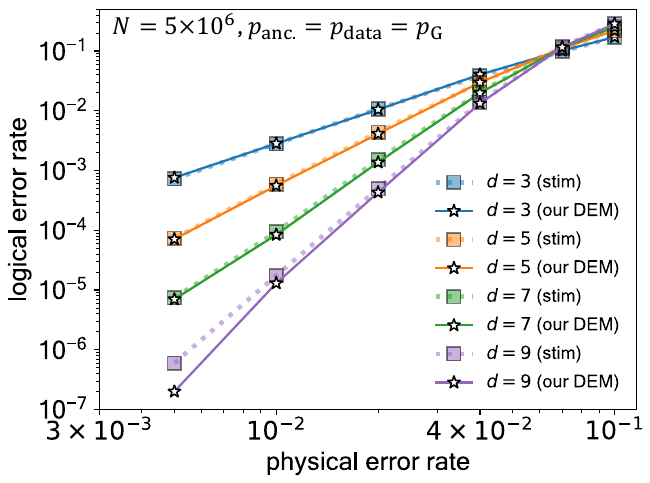}
    \caption{Logical error rate versus physical error rate for $d$ rounds of Steane-style syndrome extraction of a repetition code memory, without using a flag qubit. The dashed lines show the logical error rate extracted by the exact model of Stim for distances $d=3$ (blue), $d=5$ (orange), $d=7$ (green), and $d=9$ (purple). The solid lines show the logical error rate extracted by our estimation method. The two-qubit depolarizing rate is set equal to the single-qubit depolarizing rate of the ancilla and data qubits, which is the physical error rate. The number of shots used for the estimation is equal to the decoding shots, and set to the value $N=5\times 10^6$.    }
    \label{fig:Logical_rate_Rep_Code_Steane_No_Flag}
\end{figure}

In Fig.~\ref{fig:Logical_rate_Rep_Code_Steane_No_Flag}, we set all error rates in the circuit to be equal i.e., $p_{\text{anc}.}=p_{\text{data}}=p_\text{G}=p_{\text{phys.}}$, and vary the physical error rate. We consider distances $d\in[3,5,7]$ with $d$-rounds of syndrome extraction and sample for $5\times 10^6$ shots to do the estimation and decode via MWPM. We also set the ``decompose errors'' option of Stim's DEM and our reconstructed DEM to false. The dashed lines show Stim's performance, whereas the solid lines are obtained from our estimated DEMs. We notice that we can reconstruct the DEM and its performance with very good accuracy across all physical error rates and various distances. The threshold obtained in this case where no flag qubits are used is close to $7\%$.

\subsection{Color code with bare-ancilla syndrome extraction \label{Subsec:Color_Code}}

\begin{figure*}[!htbp]
    \centering
    \includegraphics[scale=0.655]{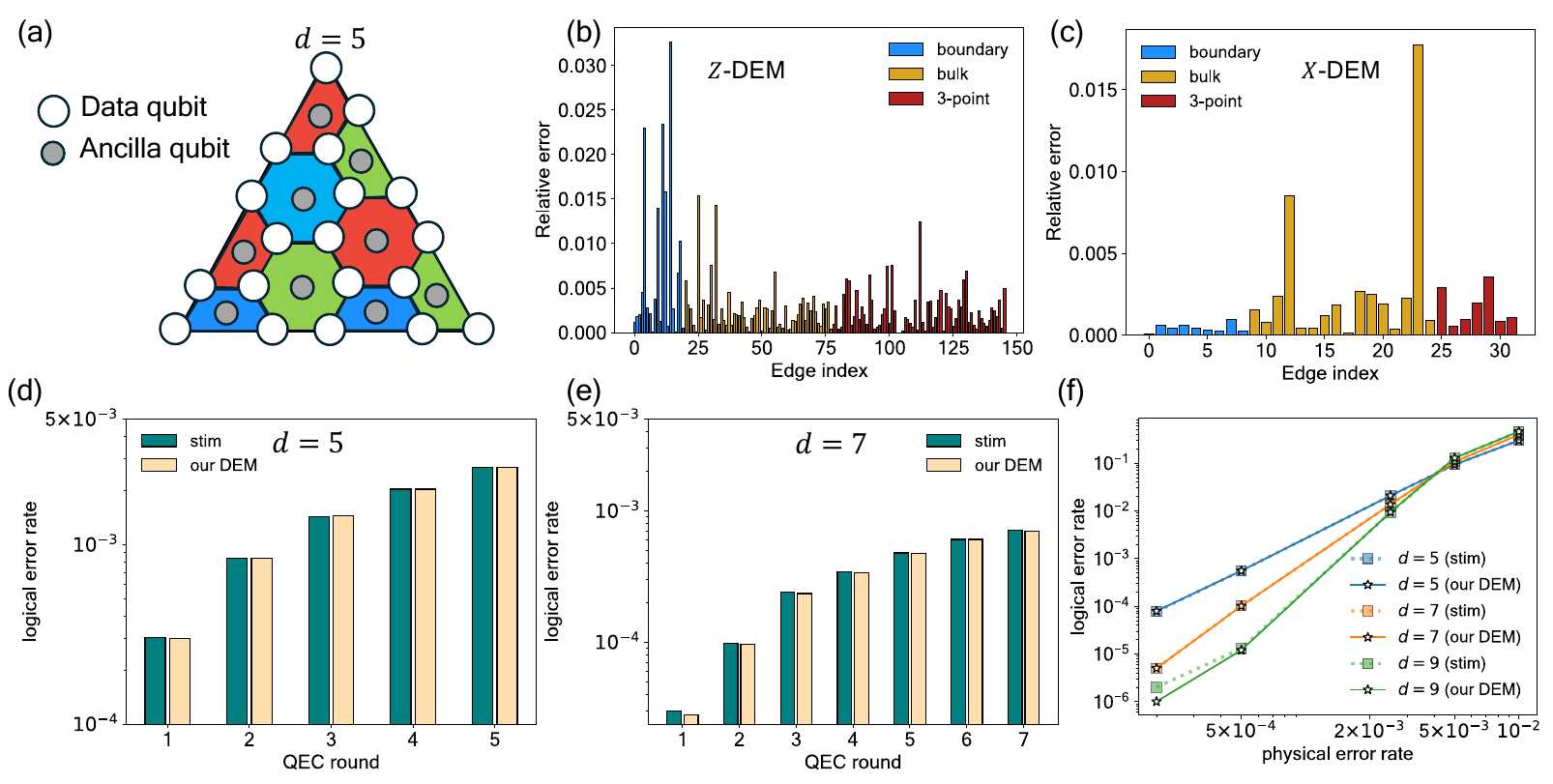}
    \caption{Estimating the decoding hypergraph of the color code memory. (a) Color code lattice for $d=5$. The white circles denote the data qubits, and the grey circles the ancilla qubits. (b) Relative error in estimating the edges (blue: boundary, yellow: bulk) and hyperedges (red: 3-point events) of the $Z$-DEM of a $d=5$ color code memory for $r=2$ QEC rounds. The error rate is set to $p=2\times 10^{-3}$, and we take $N=5\times 10^7$ shots for the estimation. (c) Same as in (b) for the $X$-DEM.  (d) Logical error rate as a function of the QEC rounds for $d=5$ color code memory, decoded with MWPM. The dark-colored bars show the error rate obtained by Stim's model and the light-colored bars show the error rate obtained by our reconstructed model. The physical error rate is set to $p=0.001$, and we use $N_{\text{est}}=10^6$ shots for the estimation, and $N_{\text{dec}}=10^6$ shots to decode. (e) Same as in (d) for $d=7$ color code memory and up to 7 QEC rounds. (f) Logical error rate as a function of the physical error rate  for $d\in [5,7,9]$ color code memory. The dashed lines correspond to Stim's DEM and the solid lines to our reconstructed DEM. We used $N=10^6$ shots for the estimation as well as the decoding.   }
    \label{fig:Color_Code}
\end{figure*}

As a following example, we apply our estimation procedure to a color code memory [see Fig.~\ref{fig:Color_Code}(a)]. The color code is promising for fault-tolerant quantum computations since all Clifford gates can be implemented transversally, a feature that also allows for resource-efficient magic state preparation~\cite{ZhangPRR2024,Bartlett2025arXiv}. Despite these advantages, the detector error model of the color code is a hypergraph, which does not allow for efficient decoding without compromising the decoding performance. There exist MWPM decoders that create smaller decoding graphs out of the entire detector error model, such as the projection decoder~\cite{DelfossePRA2014}, the Mobius decoder~\cite{KaavyaPRXQ2022,gidney2023arxiv}, or the decoder of Ref.~\cite{LeeQuantum2025}. In our work, we will use the decoder of Ref.~\cite{LeeQuantum2025}  to decode the color code and compare the performance of the noise-aware model with our reconstructed DEM. 

For completeness, let us briefly explain the decoding procedure followed in Ref~\cite{LeeQuantum2025}. Here, we will consider a $Z$-memory color code where we measure both $Z$- and $X$- checks. The first step of the decoding procedure is to separate the entire DEM into the $Z$- and $X$-DEMs. This process drops all the correlations of $Y$ errors since they are treated as independent $X$ and $Z$ errors. As a next step, each $Z$- and $X$-DEM is broken into the color-restricted and color-only lattice, which are then decoded. For example, the blue and green faces become the nodes of the red-restricted lattice. The red edges (edges that connect red faces) and the red faces are the nodes of the red-only lattice. Similar definitions hold for the other colors. MWPM is repeated for every color-restricted and color-only lattice, and the best solution is the one that corresponds to a color of minimum weight. In our work, we want to estimate the $Z$- and $X$-DEMs, so that we then separate them into the color-restricted or color-only lattices. To do that latter step, and to also finally decode the DEMs, we use the package developed by the authors of Ref.~\cite{LeeQuantum2025}, which is publicly available in Ref.~\cite{Lee2025Github}.

If we perform the noise estimation using the two-point method (and ignore higher-order correlations), then the probabilities we find reconstruct all the error events of the color-restricted lattice. This is because the error events present in this lattice flip at most two detectors, and their error probabilities include both two-point and three-point correlations. However, the two-point noise estimation method cannot reconstruct the color-only lattice, and thus, we need to resort to the multi-point correlation procedure. Similarly to the procedure explained for the repetition code with Steane-style syndrome extraction, we form equations for the configuration probabilities. The difference in this case is that we have configuration probabilities of the form $P(x_0,x_1,x_2)$. Assuming the detector triplet $D_0-D_1-D_2$ we have the following possible error events:
\begin{itemize}
    \item single-point events $p_0, p_1, p_2$
    \item two-point events $p_{01}, p_{02}, p_{12}$
    \item three-point event $p_{012}$.
\end{itemize}
We choose to keep up to $\mathcal{O}(p^7)$ terms, which actually makes the equations exact since there are 7 unknown quantities in total. We also set very loose bounds for accepting the probability solutions  obtained by the least-squares method, namely the range $[10^{-12},0.6]$.

We first consider a $d=5$ color code memory and $r=2$ QEC rounds. The circuit error model we consider has two-qubit depolarizing error, idling single-qubit depolarizing, as well as measurement,  reset and initialization errors, simulated as bit-flips or phase-flips. For simplicity, all errors are assumed to have the same strength. We solve the equations numerically for any possible triplet that exists in the actual $Z$- or $X$-DEMs obtained by Stim. Once we have the solutions of the three-point probabilities, we redefine the boundary and bulk edges of each DEM, as we did in Sec.~\ref{Subsec:Rep_Code_Steane}. The relative error in estimating the edges and hyperedges of the $Z$-DEM is shown in Fig.~\ref{fig:Color_Code}(b), and of the $X$-DEM in Fig.~\ref{fig:Color_Code}(c). The circuit error rate is set to $p=2\times 10^{-3}$, and we use $N=5\times 10^7$ shots for the estimation. We notice that the relative error for estimating the three-point probabilities is less than $1\%$, and the relative error in estimating the remaining edges is less than $4\%$.

Figure ~\ref{fig:Color_Code}(d) shows the logical error rate as a function of the QEC rounds for $d=5$. We set the circuit error rate to $p=0.001$, and use $10^6$ shots for the estimation and then the  decoding. The dark-colored bars show the performance of Stim, and the light-colored bars show the performance of our reconstructed error model. We find very good agreement in the logical error rate across all rounds. Although we use only $10^6$ shots for the estimation, which incurs a larger edge/hyperedge relative error rate than the one shown in Figs.~\ref{fig:Color_Code}(b),(c), the logical error rate accuracy is not significantly impacted. This is because the decoder is not very sensitive to the exact weights, and $10^6$ shots are still enough to reconstruct the logical error rate. A similar behavior is shown in Fig.~\ref{fig:Color_Code}(e), where we show the logical error rate for $d=7$ color code memory as a function of the QEC rounds and for the same physical error rate of $p=0.01$. Finally, in Fig.~\ref{fig:Color_Code}(f), we show the logical error rate as we vary the physical error rate for distances $d\in[5,7,9]$, for $r=d$ QEC rounds and $10^6$ shots for the estimation and decoding. The dashed lines show the performance of Stim's DEM, and the solid lines show the performance of our reconstructed DEM. Once again, we recover the logical error rate to a very good agreement with the noise-aware model, even for physical error rates exceeding the threshold.

\begin{figure}[!htbp]
    \centering
    \includegraphics[scale=0.67]{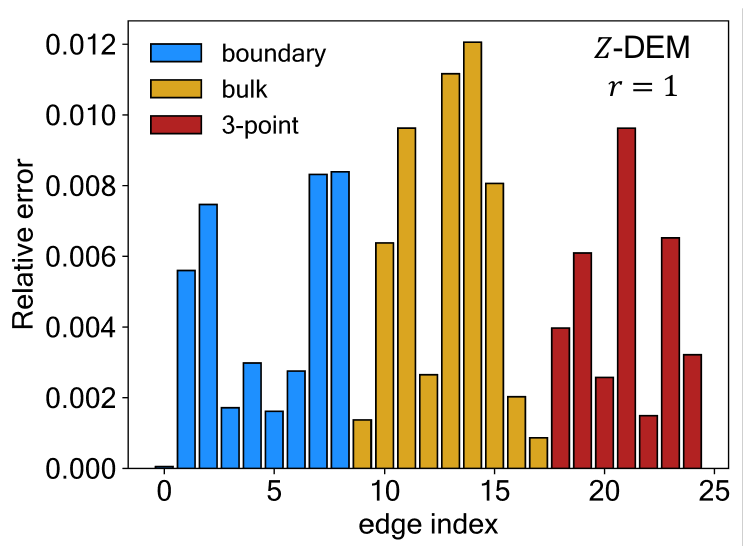}
    \caption{Relative error in estimating the edges of a $d=3$ color code memory, for $r=1$ QEC round. The error rate of the circuit is set to $p=10^{-3}$, and we use $N_{\text{est}}=2\times 10^7$ shots for the estimation. The blue bars show the relative error for boundary edges, the yellow bars for bulk edges, and the red bars for the hyperdeges.  }
    \label{fig:Color_code_d_3}
\end{figure}
At this point, the reader might notice that, so far, we have shown results for the color code only for distances $d\geq 5$. This is because the DEM of a $d=3$ color code memory exhibits an interesting feature that we analyze now separately. Performing the analysis we mentioned for $d\geq 5$,  allows us to obtain almost all error probabilities present in the DEM of a $d=3$ color code. However, there exist errors, which can cause a single detector to fire, and events, which can flip the same detector and the logical observable. In other words, for certain detectors, we can have the error events $D_j$ and $D_j-L_0$. If we assume the first event happens with probability $p$ and the latter event happens with probability $q$, then our estimation procedure for a boundary node finds the total probability $\alpha=p+q-2pq$. 

In order to distinguish the error probabilities of $D_j$ and $D_j-L_0$, for any possible $j$ where this happens, we introduce a new system of equations that we solve numerically. We form the configuration probability equation for $P(D_j-L_0)$ as follows. We collect all possible error mechanisms that exist in the DEM, and we create terms that flip $D_j$ as well as the logical observable $L_0$ an odd number of times, whereas they flip all other detectors an even number of times. We further substitute the expression $p=(\alpha-q)/(1-2q)$ since we know $\alpha$. Then, we count from statistics the occurrence of such patterns and set $P(D_j-L_0)$ to the value obtained from statistics. The equations we form are coupled, but we can solve them using the least-squares method. 

We should comment that not every detector in the DEM breaks into these two errors $D_j$ and $D_j-L_0$, so the equations we need to solve are less than the number of detectors $D_j$. Further, as the number of QEC rounds increases, so does the size of the numerical system and the number of detectors we need to include in the equations. Nevertheless, we can truncate to relatively low order in $p$ and estimate with good accuracy the unknown probabilities.

To demonstrate how well this numerical solution performs, we apply it to the case of $r=1$ QEC round and truncate the numerical equations to $\mathcal{O}(p^2)$.
For $r=1$ QEC round, we only have access to the $Z$-DEM, since the first round of $X$-type measurements is random.
We use $N=2\times 10^7$ shots for the estimation and set the error rate of the circuit to $p=10^{-3}$. In Fig.~\ref{fig:Color_code_d_3}, we show the relative error in estimating the edges and hyperedges of the $Z$-DEM. We note excellent agreement across all types of edges and hyperedges, with a relative error within $\sim 1.2\%$. The solution proposed here might appear in other QEC codes and their DEMs, so it is a general method of constructing equations to separate contributions of events that flip a set of detectors, from events that flip a set of detectors together with the logical observable. 

\section{Conclusions \label{Sec:Conclusions}}
Estimating the noise sources of quantum devices is essential for scaling up error-corrected architectures. Further, noise-aware decoding provides useful information to the decoder which can lead to improvements in the decoder's performance. We showed how to reconstruct decoding graphs or hypergraphs of various QEC codes, by using only the syndrome history. We compared the logical error rate of the noise-aware models and of our reconstructed error models and we find excellent agreement between them. For the case of hypergraphs, we can analyze a few-point regions of the hypergraph to extract the error rates of higher-order events, and then redefine the remaining edge probabilities. Our methods can be adapted to learn Pauli noise of other QEC codes and experiments beyond memory experiments, given that typically higher-order correlations are bounded by some small constant. When finalizing our manuscript we encountered Ref.~\cite{WallraffArXiv2025} which also
studies correlations in the syndrome statistics and found closed-form expressions for the higher-order probabilities. Their method is applied to a $d=3$ rotated surface code. We tested their closed-form expression for three-point error probabilities on the color-code memory and we find that our numerical solutions match to very high precision with the analytical solutions. We further tested their four-point probability formula on the repetition code under Steane-style syndrome extraction, and this also matches with our numerical solutions. 

{\color{black}{Related to ACES, the method developed in Ref.~\cite{HarperPRXQ2025} is involved and requires optimization of tuples corresponding to the layers of a quantum circuit which is exponentially hard for a full search, and not feasible for large code distances. However, the authors deal with this issue by employing a greedy selection, and they only 
optimize the tuples sets for small code distances which can then be used for larger code distances. Further, the Pauli channel probabilities they extract from their estimation method need to be projected onto the probability simplex to be physically valid~\cite{FlamiaACES2022,Harper2025arXiv}, which also incurs an extra computational cost. To estimate $G_1$ single-qubit gate eigenvalues and $G_2$ two-qubit gate eigenvalues for $N$ qubit-circuits up to fidelity $\epsilon$, $\mathcal{O}(N(G_1+G_2)\epsilon^{-2})$ measurements are needed~\cite{FlamiaACES2022,PelaezArXiv2024}, so  Refs.~\cite{Harper2025arXiv, HarperPRXQ2025} can obtain very good accuracy in the estimation with a feasible number of experimental shots. For a decoding graph such as the surface code, where the number of qubits scales as $\mathcal{O}(d^2)$, we need to estimate two-point correlations between nearest-neighbor detectors, leading to a classical post-processing scaling $\mathcal{O}(m\times c\times r)$, where $m$ is the number of detectors ($\mathcal{O}(d^2)$ ancilla), $r\sim \mathcal{O}(d)$ is the number of QEC rounds, and $c$ is the number of nearest detectors to a given detector. For a fixed small number of rounds, our noise estimation method achieves a $\mathcal{O}(n)\sim \mathcal{O}(d^2)$ computational cost for reconstructing the DEM of a surface code. This scaling could be transferable to other quantum low-density parity check (Q-LDPC) codes. ACES achieves a scaling of $\mathcal{O}(n)$, but this provides only an average error rate across rounds, whereas in our methods the error rates can differ across rounds.}}

\section{Outlook}
An interesting future direction would be to study the effect of noise drift in the estimation and introduce other sources of noise such as leakage, cross-talk or qubit loss. Another avenue would be to apply the noise estimation method to coherent noise models, identify regimes where the Pauli approximation is consistent, and explore potential differences in statistics between coherent and stochastic Pauli noise.

\acknowledgements{The authors acknowledge support from  the Office of the Director of National Intelligence (ODNI), Intelligence Advanced Research Projects Activity (IARPA), under the Entangled Logical Qubits program through Cooperative Agreement Number W911NF-23-2-0216 and the ARO/LPS QCISS program (W911NF-21-1-0005).}

% \nocite{*}

\appendix

\section{Limit of two-qubit errors for bare-ancilla repetition code memory \label{App:Two_qubit_error_limit_rep_code_bare_ancilla}}

\begin{figure}[!htbp]
    \centering
    \includegraphics[scale=0.88]{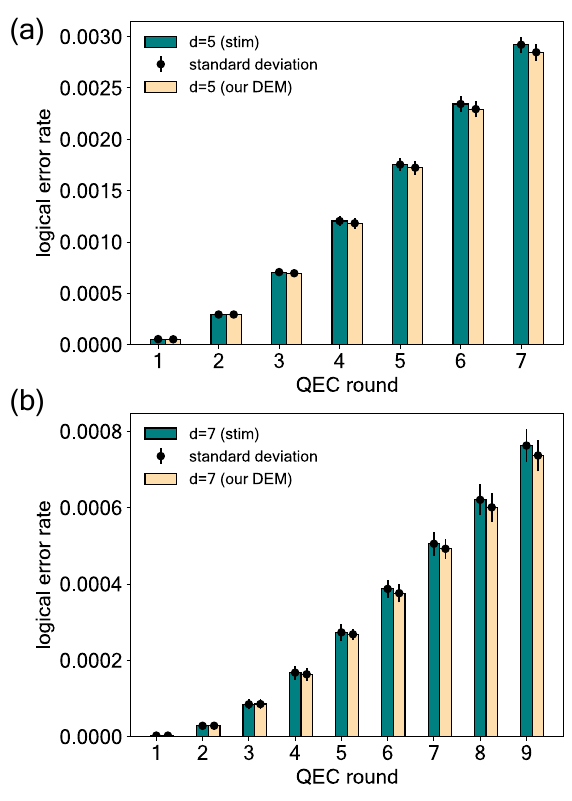}
    \caption{Average logical error rate versus QEC rounds for $d=5$ (a) and $d=7$ (b) repetition code memory under bare-ancilla syndrome extraction. The gate error is set to $p_G=0.025$, and the qubit errors are assumed to be zero. The error bars show the standard deviation obtained by repeating the sampling for 50 times. }
    \label{fig:App_Rep_Code}
\end{figure}

Here we study the limiting case of two-qubit gate errors for the repetition code memory, with bare-ancilla syndrome extraction. We consider the gate error of $p_G=0.025$ and sample for $N=5\times 10^5$ shots to estimate and then decode both our DEM and Stim's DEM. We repeat this process 50 times to collect the average logical error and the standard deviation. Our results are shown in Fig.~\ref{fig:App_Rep_Code}(a) for $d=5$, and in Fig.~\ref{fig:App_Rep_Code}(b) for $d=7$. We note that again our estimated DEM outperforms Stim's DEM as the number of QEC rounds increases. The improvement in the logical error rate is less pronounced compared to the one we showed in the main text for $d=3$. It is also apparent from Fig.~\ref{fig:App_Rep_Code} that we need to sample for more shots to reduce the standard deviation as we increase the distance and the QEC rounds. Overal, we see that the average logical error  of our DEM across all QEC rounds is smaller than (or equal to) Stim's logical error rate.

\section{Improvement for $d=5$ surface code under varying error rates \label{App:Surface_Code_Varying_d_5}}
\begin{figure}
    \centering
    \includegraphics[scale=0.8]{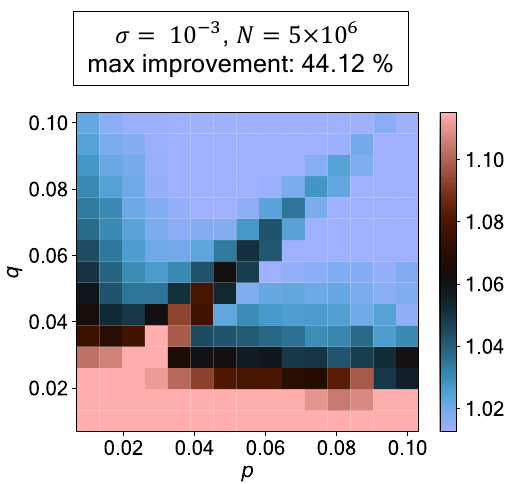}
    \caption{Comparison of logical error rate,  $p_L^{(\text{stim})}$, obtained based on fixed error probabilities $p$ on data qubits and fixed error probabilities $q$ on ancilla qubits, with the logical error rate $p_L^{(\text{our DEM})}$ of a model where the error rates fluctuate across qubits, for a distance $d=5$ surface code memory and $r=5$ QEC rounds. The error rates are sampled from the log-normal distribution with a standard deviation $\sigma$. We consider only the $X$-type stabilizers and decode the $X$-DEM. Our noise model is input single-qubit depolarizing rate on both data and ancilla qubits, and we assume perfect gates. The color shows the ratio of $p_L^{(\text{stim})}/p_L^{(\text{our DEM})}$. The number of shots used for the estimation and the decoding is $N=5\times 10^6$, and the standard deviation for the log-normal distribution is $\sigma = 10^{-3}$. The maximum improvement we find is $44.12\%$. The colorbar range is scaled to the same minimum and maximum value as in Fig.~\ref{fig:Surface_Code_Phenom}.}
    \label{fig:Surface_Code_Phenom_d_5}
\end{figure}

In this section we present more results for the case of varying error rates on a surface code memory. Similar to the simulation presented in Sec.~\ref{Subsec:Two_point}, we assume a $d=5$ surface code $X$-memory, where syndrome extraction is repeated for $r=5$ QEC rounds. We only measure $Z$-type stabilizers, assume perfect gates, and input single-qubit depolarizing error on data and ancilla qubits. We consider the probability range $p,q\in[0.01,0.1]$ for the mean values of error probabilities for data and ancilla qubits. The error probabilities are again sampled from the log-normal probability distribution, and we set the standard deviation to 
$\sigma=10^{-3}$. Further, we use $N=5\times 10^6$ shots for the estimation and decoding of each DEM. In Fig.~\ref{fig:Surface_Code_Phenom_d_5} we show with color the ratio of stim's logical error rate (where the standard deviation for physical error rates is set to 0), compared to the logical error rate we obtained from the error model where physical error rates vary across the qubits. We find an even bigger improvement range compared to $d=3$ surface code, with a maximum improvement of $44.12\%$.

\section{Runtime of estimation for unrotated surface code}
\begin{figure}[!htbp]
    \centering
    \includegraphics[scale=0.7]{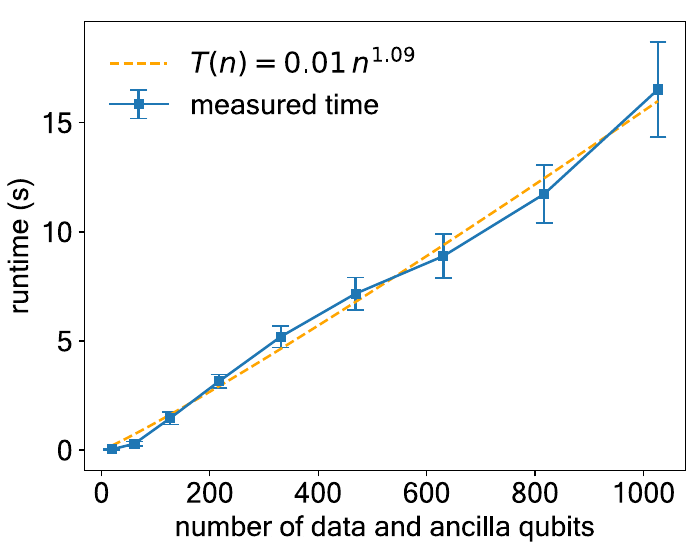}
    \caption{Average runtime to estimate the $X$-DEM of the unrotated surface code, as a function of the total number of qubits. The minimum distance is $d=3$, and the maximum distance is $d=19$. We also fix the number of QEC rounds to $r=3$. To collect the average runtime, we repeat the calculation $M=400$ times. The blue line shows the elapsed time we measure, with the error bars displaying the standard deviation, and the dashed line is a fit. }
    \label{fig:Runtime_of_Est_Surface_Code}
\end{figure}

Here we test the runtime for extracting the error probabilities of a surface code DEM. In particular we consider the unrotated surface code which consists of $L^2+(L-1)^2$ data qubits, and $L(L-1)$ ancilla qubits (for one type of checks), and we assume that we measure the $X$-type stabilizers. We assume distances $d\in[3,19]$ with a step of 2, and a fixed number of QEC rounds, $r=3$. The total number of shots we use for the estimation is set to $N=10^5$. We calculate the runtime after we obtain the defects matrix. Specifically, the runtime is associated to the time it takes us to estimate the time and bulk edges of the DEM, given a defects matrix of $N=10^5$ shots. To collect an average runtime per distance, we repeat the calculation $M=400$ times. Figure~\ref{fig:Runtime_of_Est_Surface_Code} shows the average runtime as a function of the total number of qubits of the unrotated surface code. We fit an exponential function with respect to the number of qubits, and we find a scaling slightly higher than $\mathcal{O}(n)$.  This verifies that for a fixed small number of rounds, the scaling of our estimation for the unrotated surface code is close to $\mathcal{O}(n)$.

\clearpage


\begin{thebibliography}{59}%
\makeatletter
\providecommand \@ifxundefined [1]{%
 \@ifx{#1\undefined}
}%
\providecommand \@ifnum [1]{%
 \ifnum #1\expandafter \@firstoftwo
 \else \expandafter \@secondoftwo
 \fi
}%
\providecommand \@ifx [1]{%
 \ifx #1\expandafter \@firstoftwo
 \else \expandafter \@secondoftwo
 \fi
}%
\providecommand \natexlab [1]{#1}%
\providecommand \enquote  [1]{``#1''}%
\providecommand \bibnamefont  [1]{#1}%
\providecommand \bibfnamefont [1]{#1}%
\providecommand \citenamefont [1]{#1}%
\providecommand \href@noop [0]{\@secondoftwo}%
\providecommand \href [0]{\begingroup \@sanitize@url \@href}%
\providecommand \@href[1]{\@@startlink{#1}\@@href}%
\providecommand \@@href[1]{\endgroup#1\@@endlink}%
\providecommand \@sanitize@url [0]{\catcode `\\12\catcode `\$12\catcode `\&12\catcode `\#12\catcode `\^12\catcode `\_12\catcode `\%12\relax}%
\providecommand \@@startlink[1]{}%
\providecommand \@@endlink[0]{}%
\providecommand \url  [0]{\begingroup\@sanitize@url \@url }%
\providecommand \@url [1]{\endgroup\@href {#1}{\urlprefix }}%
\providecommand \urlprefix  [0]{URL }%
\providecommand \Eprint [0]{\href }%
\providecommand \doibase [0]{https://doi.org/}%
\providecommand \selectlanguage [0]{\@gobble}%
\providecommand \bibinfo  [0]{\@secondoftwo}%
\providecommand \bibfield  [0]{\@secondoftwo}%
\providecommand \translation [1]{[#1]}%
\providecommand \BibitemOpen [0]{}%
\providecommand \bibitemStop [0]{}%
\providecommand \bibitemNoStop [0]{.\EOS\space}%
\providecommand \EOS [0]{\spacefactor3000\relax}%
\providecommand \BibitemShut  [1]{\csname bibitem#1\endcsname}%
\let\auto@bib@innerbib\@empty
%</preamble>
\bibitem [{\citenamefont {AI}\ and\ \citenamefont {Collaborators}(2025)}]{GoogleNature2025}%
  \BibitemOpen
  \bibfield  {author} {\bibinfo {author} {\bibfnamefont {G.~Q.}\ \bibnamefont {AI}}\ and\ \bibinfo {author} {\bibnamefont {Collaborators}},\ }\bibfield  {title} {\bibinfo {title} {Quantum error correction below the surface code threshold},\ }\href {https://doi.org/10.1038/s41586-024-08449-y} {\bibfield  {journal} {\bibinfo  {journal} {Nature}\ }\textbf {\bibinfo {volume} {638}},\ \bibinfo {pages} {920–926} (\bibinfo {year} {2025})}\BibitemShut {NoStop}%
\bibitem [{\citenamefont {Erhard}\ \emph {et~al.}(2021)\citenamefont {Erhard}, \citenamefont {Nautrup}, \citenamefont {Meth}, \citenamefont {Postler}, \citenamefont {Stricker}, \citenamefont {Stadler}, \citenamefont {Negnevitsky}, \citenamefont {Ringbauer}, \citenamefont {Schindler}, \citenamefont {Briegel}, \citenamefont {Blatt}, \citenamefont {Friis},\ and\ \citenamefont {Monz}}]{ErhardNature2021}%
  \BibitemOpen
  \bibfield  {author} {\bibinfo {author} {\bibfnamefont {A.}~\bibnamefont {Erhard}}, \bibinfo {author} {\bibfnamefont {H.~P.}\ \bibnamefont {Nautrup}}, \bibinfo {author} {\bibfnamefont {M.}~\bibnamefont {Meth}}, \bibinfo {author} {\bibfnamefont {L.}~\bibnamefont {Postler}}, \bibinfo {author} {\bibfnamefont {R.}~\bibnamefont {Stricker}}, \bibinfo {author} {\bibfnamefont {M.}~\bibnamefont {Stadler}}, \bibinfo {author} {\bibfnamefont {V.}~\bibnamefont {Negnevitsky}}, \bibinfo {author} {\bibfnamefont {M.}~\bibnamefont {Ringbauer}}, \bibinfo {author} {\bibfnamefont {P.}~\bibnamefont {Schindler}}, \bibinfo {author} {\bibfnamefont {H.~J.}\ \bibnamefont {Briegel}}, \bibinfo {author} {\bibfnamefont {R.}~\bibnamefont {Blatt}}, \bibinfo {author} {\bibfnamefont {N.}~\bibnamefont {Friis}},\ and\ \bibinfo {author} {\bibfnamefont {T.}~\bibnamefont {Monz}},\ }\bibfield  {title} {\bibinfo {title} {Entangling logical qubits with lattice surgery},\ }\href {https://doi.org/10.1038/s41586-020-03079-6} {\bibfield  {journal}
  {\bibinfo  {journal} {Nature}\ }\textbf {\bibinfo {volume} {589}},\ \bibinfo {pages} {220–224} (\bibinfo {year} {2021})}\BibitemShut {NoStop}%
\bibitem [{\citenamefont {Ryan-Anderson}\ \emph {et~al.}(2024)\citenamefont {Ryan-Anderson}, \citenamefont {Brown}, \citenamefont {Baldwin}, \citenamefont {Dreiling}, \citenamefont {Foltz}, \citenamefont {Gaebler}, \citenamefont {Gatterman}, \citenamefont {Hewitt}, \citenamefont {Holliman}, \citenamefont {Horst}, \citenamefont {Johansen}, \citenamefont {Lucchetti}, \citenamefont {Mengle}, \citenamefont {Matheny}, \citenamefont {Matsuoka}, \citenamefont {Mayer}, \citenamefont {Mills}, \citenamefont {Moses}, \citenamefont {Neyenhuis}, \citenamefont {Pino}, \citenamefont {Siegfried}, \citenamefont {Stutz}, \citenamefont {Walker},\ and\ \citenamefont {Hayes}}]{Anderson2024arXiv}%
  \BibitemOpen
  \bibfield  {author} {\bibinfo {author} {\bibfnamefont {C.}~\bibnamefont {Ryan-Anderson}}, \bibinfo {author} {\bibfnamefont {N.~C.}\ \bibnamefont {Brown}}, \bibinfo {author} {\bibfnamefont {C.~H.}\ \bibnamefont {Baldwin}}, \bibinfo {author} {\bibfnamefont {J.~M.}\ \bibnamefont {Dreiling}}, \bibinfo {author} {\bibfnamefont {C.}~\bibnamefont {Foltz}}, \bibinfo {author} {\bibfnamefont {J.~P.}\ \bibnamefont {Gaebler}}, \bibinfo {author} {\bibfnamefont {T.~M.}\ \bibnamefont {Gatterman}}, \bibinfo {author} {\bibfnamefont {N.}~\bibnamefont {Hewitt}}, \bibinfo {author} {\bibfnamefont {C.}~\bibnamefont {Holliman}}, \bibinfo {author} {\bibfnamefont {C.~V.}\ \bibnamefont {Horst}}, \bibinfo {author} {\bibfnamefont {J.}~\bibnamefont {Johansen}}, \bibinfo {author} {\bibfnamefont {D.}~\bibnamefont {Lucchetti}}, \bibinfo {author} {\bibfnamefont {T.}~\bibnamefont {Mengle}}, \bibinfo {author} {\bibfnamefont {M.}~\bibnamefont {Matheny}}, \bibinfo {author} {\bibfnamefont {Y.}~\bibnamefont {Matsuoka}}, \bibinfo {author}
  {\bibfnamefont {K.}~\bibnamefont {Mayer}}, \bibinfo {author} {\bibfnamefont {M.}~\bibnamefont {Mills}}, \bibinfo {author} {\bibfnamefont {S.~A.}\ \bibnamefont {Moses}}, \bibinfo {author} {\bibfnamefont {B.}~\bibnamefont {Neyenhuis}}, \bibinfo {author} {\bibfnamefont {J.}~\bibnamefont {Pino}}, \bibinfo {author} {\bibfnamefont {P.}~\bibnamefont {Siegfried}}, \bibinfo {author} {\bibfnamefont {R.~P.}\ \bibnamefont {Stutz}}, \bibinfo {author} {\bibfnamefont {J.}~\bibnamefont {Walker}},\ and\ \bibinfo {author} {\bibfnamefont {D.}~\bibnamefont {Hayes}},\ }\href {https://arxiv.org/abs/2404.16728} {\bibinfo {title} {High-fidelity and fault-tolerant teleportation of a logical qubit using transversal gates and lattice surgery on a trapped-ion quantum computer}} (\bibinfo {year} {2024}),\ \Eprint {https://arxiv.org/abs/2404.16728} {arXiv:2404.16728 [quant-ph]} \BibitemShut {NoStop}%
\bibitem [{\citenamefont {AI}\ and\ \citenamefont {Collaborators}(2024{\natexlab{a}})}]{LacroixArXiv2024}%
  \BibitemOpen
  \bibfield  {author} {\bibinfo {author} {\bibfnamefont {G.~Q.}\ \bibnamefont {AI}}\ and\ \bibinfo {author} {\bibnamefont {Collaborators}},\ }\href {https://arxiv.org/abs/2412.14256} {\bibinfo {title} {Scaling and logic in the color code on a superconducting quantum processor}} (\bibinfo {year} {2024}{\natexlab{a}}),\ \Eprint {https://arxiv.org/abs/2412.14256} {arXiv:2412.14256 [quant-ph]} \BibitemShut {NoStop}%
\bibitem [{\citenamefont {Rodriguez}\ \emph {et~al.}(2024)\citenamefont {Rodriguez}, \citenamefont {Robinson}, \citenamefont {Jepsen}, \citenamefont {He}, \citenamefont {Duckering}, \citenamefont {Zhao}, \citenamefont {Wu}, \citenamefont {Campo}, \citenamefont {Bagnall}, \citenamefont {Kwon}, \citenamefont {Karolyshyn}, \citenamefont {Weinberg}, \citenamefont {Cain}, \citenamefont {Evered}, \citenamefont {Geim}, \citenamefont {Kalinowski}, \citenamefont {Li}, \citenamefont {Manovitz}, \citenamefont {Amato-Grill}, \citenamefont {Basham}, \citenamefont {Bernstein}, \citenamefont {Braverman}, \citenamefont {Bylinskii}, \citenamefont {Choukri}, \citenamefont {DeAngelo}, \citenamefont {Fang}, \citenamefont {Fieweger}, \citenamefont {Frederick}, \citenamefont {Haines}, \citenamefont {Hamdan}, \citenamefont {Hammett}, \citenamefont {Hsu}, \citenamefont {Hu}, \citenamefont {Huber}, \citenamefont {Jia}, \citenamefont {Kedar}, \citenamefont {Kornjača}, \citenamefont {Liu}, \citenamefont {Long}, \citenamefont
  {Lopatin}, \citenamefont {Lopes}, \citenamefont {Luo}, \citenamefont {Macrì}, \citenamefont {Marković}, \citenamefont {Martínez-Martínez}, \citenamefont {Meng}, \citenamefont {Ostermann}, \citenamefont {Ostroumov}, \citenamefont {Paquette}, \citenamefont {Qiang}, \citenamefont {Shofman}, \citenamefont {Singh}, \citenamefont {Singh}, \citenamefont {Sinha}, \citenamefont {Thoreen}, \citenamefont {Wan}, \citenamefont {Wang}, \citenamefont {Waxman-Lenz}, \citenamefont {Wong}, \citenamefont {Wurtz}, \citenamefont {Zhdanov}, \citenamefont {Zheng}, \citenamefont {Greiner}, \citenamefont {Keesling}, \citenamefont {Gemelke}, \citenamefont {Vuletić}, \citenamefont {Kitagawa}, \citenamefont {Wang}, \citenamefont {Bluvstein}, \citenamefont {Lukin}, \citenamefont {Lukin}, \citenamefont {Zhou},\ and\ \citenamefont {Cantú}}]{Rodriguez2024arXiv}%
  \BibitemOpen
  \bibfield  {author} {\bibinfo {author} {\bibfnamefont {P.~S.}\ \bibnamefont {Rodriguez}}, \bibinfo {author} {\bibfnamefont {J.~M.}\ \bibnamefont {Robinson}}, \bibinfo {author} {\bibfnamefont {P.~N.}\ \bibnamefont {Jepsen}}, \bibinfo {author} {\bibfnamefont {Z.}~\bibnamefont {He}}, \bibinfo {author} {\bibfnamefont {C.}~\bibnamefont {Duckering}}, \bibinfo {author} {\bibfnamefont {C.}~\bibnamefont {Zhao}}, \bibinfo {author} {\bibfnamefont {K.-H.}\ \bibnamefont {Wu}}, \bibinfo {author} {\bibfnamefont {J.}~\bibnamefont {Campo}}, \bibinfo {author} {\bibfnamefont {K.}~\bibnamefont {Bagnall}}, \bibinfo {author} {\bibfnamefont {M.}~\bibnamefont {Kwon}}, \bibinfo {author} {\bibfnamefont {T.}~\bibnamefont {Karolyshyn}}, \bibinfo {author} {\bibfnamefont {P.}~\bibnamefont {Weinberg}}, \bibinfo {author} {\bibfnamefont {M.}~\bibnamefont {Cain}}, \bibinfo {author} {\bibfnamefont {S.~J.}\ \bibnamefont {Evered}}, \bibinfo {author} {\bibfnamefont {A.~A.}\ \bibnamefont {Geim}}, \bibinfo {author} {\bibfnamefont
  {M.}~\bibnamefont {Kalinowski}}, \bibinfo {author} {\bibfnamefont {S.~H.}\ \bibnamefont {Li}}, \bibinfo {author} {\bibfnamefont {T.}~\bibnamefont {Manovitz}}, \bibinfo {author} {\bibfnamefont {J.}~\bibnamefont {Amato-Grill}}, \bibinfo {author} {\bibfnamefont {J.~I.}\ \bibnamefont {Basham}}, \bibinfo {author} {\bibfnamefont {L.}~\bibnamefont {Bernstein}}, \bibinfo {author} {\bibfnamefont {B.}~\bibnamefont {Braverman}}, \bibinfo {author} {\bibfnamefont {A.}~\bibnamefont {Bylinskii}}, \bibinfo {author} {\bibfnamefont {A.}~\bibnamefont {Choukri}}, \bibinfo {author} {\bibfnamefont {R.}~\bibnamefont {DeAngelo}}, \bibinfo {author} {\bibfnamefont {F.}~\bibnamefont {Fang}}, \bibinfo {author} {\bibfnamefont {C.}~\bibnamefont {Fieweger}}, \bibinfo {author} {\bibfnamefont {P.}~\bibnamefont {Frederick}}, \bibinfo {author} {\bibfnamefont {D.}~\bibnamefont {Haines}}, \bibinfo {author} {\bibfnamefont {M.}~\bibnamefont {Hamdan}}, \bibinfo {author} {\bibfnamefont {J.}~\bibnamefont {Hammett}}, \bibinfo {author} {\bibfnamefont
  {N.}~\bibnamefont {Hsu}}, \bibinfo {author} {\bibfnamefont {M.-G.}\ \bibnamefont {Hu}}, \bibinfo {author} {\bibfnamefont {F.}~\bibnamefont {Huber}}, \bibinfo {author} {\bibfnamefont {N.}~\bibnamefont {Jia}}, \bibinfo {author} {\bibfnamefont {D.}~\bibnamefont {Kedar}}, \bibinfo {author} {\bibfnamefont {M.}~\bibnamefont {Kornjača}}, \bibinfo {author} {\bibfnamefont {F.}~\bibnamefont {Liu}}, \bibinfo {author} {\bibfnamefont {J.}~\bibnamefont {Long}}, \bibinfo {author} {\bibfnamefont {J.}~\bibnamefont {Lopatin}}, \bibinfo {author} {\bibfnamefont {P.~L.~S.}\ \bibnamefont {Lopes}}, \bibinfo {author} {\bibfnamefont {X.-Z.}\ \bibnamefont {Luo}}, \bibinfo {author} {\bibfnamefont {T.}~\bibnamefont {Macrì}}, \bibinfo {author} {\bibfnamefont {O.}~\bibnamefont {Marković}}, \bibinfo {author} {\bibfnamefont {L.~A.}\ \bibnamefont {Martínez-Martínez}}, \bibinfo {author} {\bibfnamefont {X.}~\bibnamefont {Meng}}, \bibinfo {author} {\bibfnamefont {S.}~\bibnamefont {Ostermann}}, \bibinfo {author} {\bibfnamefont
  {E.}~\bibnamefont {Ostroumov}}, \bibinfo {author} {\bibfnamefont {D.}~\bibnamefont {Paquette}}, \bibinfo {author} {\bibfnamefont {Z.}~\bibnamefont {Qiang}}, \bibinfo {author} {\bibfnamefont {V.}~\bibnamefont {Shofman}}, \bibinfo {author} {\bibfnamefont {A.}~\bibnamefont {Singh}}, \bibinfo {author} {\bibfnamefont {M.}~\bibnamefont {Singh}}, \bibinfo {author} {\bibfnamefont {N.}~\bibnamefont {Sinha}}, \bibinfo {author} {\bibfnamefont {H.}~\bibnamefont {Thoreen}}, \bibinfo {author} {\bibfnamefont {N.}~\bibnamefont {Wan}}, \bibinfo {author} {\bibfnamefont {Y.}~\bibnamefont {Wang}}, \bibinfo {author} {\bibfnamefont {D.}~\bibnamefont {Waxman-Lenz}}, \bibinfo {author} {\bibfnamefont {T.}~\bibnamefont {Wong}}, \bibinfo {author} {\bibfnamefont {J.}~\bibnamefont {Wurtz}}, \bibinfo {author} {\bibfnamefont {A.}~\bibnamefont {Zhdanov}}, \bibinfo {author} {\bibfnamefont {L.}~\bibnamefont {Zheng}}, \bibinfo {author} {\bibfnamefont {M.}~\bibnamefont {Greiner}}, \bibinfo {author} {\bibfnamefont {A.}~\bibnamefont
  {Keesling}}, \bibinfo {author} {\bibfnamefont {N.}~\bibnamefont {Gemelke}}, \bibinfo {author} {\bibfnamefont {V.}~\bibnamefont {Vuletić}}, \bibinfo {author} {\bibfnamefont {T.}~\bibnamefont {Kitagawa}}, \bibinfo {author} {\bibfnamefont {S.-T.}\ \bibnamefont {Wang}}, \bibinfo {author} {\bibfnamefont {D.}~\bibnamefont {Bluvstein}}, \bibinfo {author} {\bibfnamefont {M.~D.}\ \bibnamefont {Lukin}}, \bibinfo {author} {\bibfnamefont {A.}~\bibnamefont {Lukin}}, \bibinfo {author} {\bibfnamefont {H.}~\bibnamefont {Zhou}},\ and\ \bibinfo {author} {\bibfnamefont {S.~H.}\ \bibnamefont {Cantú}},\ }\href {https://arxiv.org/abs/2412.15165} {\bibinfo {title} {Experimental demonstration of logical magic state distillation}} (\bibinfo {year} {2024}),\ \Eprint {https://arxiv.org/abs/2412.15165} {arXiv:2412.15165 [quant-ph]} \BibitemShut {NoStop}%
\bibitem [{\citenamefont {AI}\ and\ \citenamefont {Collaborators}(2024{\natexlab{b}})}]{EickbuschArXiv2024}%
  \BibitemOpen
  \bibfield  {author} {\bibinfo {author} {\bibfnamefont {G.~Q.}\ \bibnamefont {AI}}\ and\ \bibinfo {author} {\bibnamefont {Collaborators}},\ }\href {https://arxiv.org/abs/2412.14360} {\bibinfo {title} {Demonstrating dynamic surface codes}} (\bibinfo {year} {2024}{\natexlab{b}}),\ \Eprint {https://arxiv.org/abs/2412.14360} {arXiv:2412.14360 [quant-ph]} \BibitemShut {NoStop}%
\bibitem [{\citenamefont {Chuang}\ and\ \citenamefont {and}(1997)}]{Chuang01111997}%
  \BibitemOpen
  \bibfield  {author} {\bibinfo {author} {\bibfnamefont {I.~L.}\ \bibnamefont {Chuang}}\ and\ \bibinfo {author} {\bibfnamefont {M.~A.~N.}\ \bibnamefont {and}},\ }\bibfield  {title} {\bibinfo {title} {Prescription for experimental determination of the dynamics of a quantum black box},\ }\href {https://doi.org/10.1080/09500349708231894} {\bibfield  {journal} {\bibinfo  {journal} {Journal of Modern Optics}\ }\textbf {\bibinfo {volume} {44}},\ \bibinfo {pages} {2455} (\bibinfo {year} {1997})}\BibitemShut {NoStop}%
\bibitem [{\citenamefont {Cramer}\ \emph {et~al.}(2010)\citenamefont {Cramer}, \citenamefont {Plenio}, \citenamefont {Flammia}, \citenamefont {Somma}, \citenamefont {Gross}, \citenamefont {Bartlett}, \citenamefont {Landon-Cardinal}, \citenamefont {Poulin},\ and\ \citenamefont {Liu}}]{CramerNatCommun2010}%
  \BibitemOpen
  \bibfield  {author} {\bibinfo {author} {\bibfnamefont {M.}~\bibnamefont {Cramer}}, \bibinfo {author} {\bibfnamefont {M.~B.}\ \bibnamefont {Plenio}}, \bibinfo {author} {\bibfnamefont {S.~T.}\ \bibnamefont {Flammia}}, \bibinfo {author} {\bibfnamefont {R.}~\bibnamefont {Somma}}, \bibinfo {author} {\bibfnamefont {D.}~\bibnamefont {Gross}}, \bibinfo {author} {\bibfnamefont {S.~D.}\ \bibnamefont {Bartlett}}, \bibinfo {author} {\bibfnamefont {O.}~\bibnamefont {Landon-Cardinal}}, \bibinfo {author} {\bibfnamefont {D.}~\bibnamefont {Poulin}},\ and\ \bibinfo {author} {\bibfnamefont {Y.-K.}\ \bibnamefont {Liu}},\ }\bibfield  {title} {\bibinfo {title} {Efficient quantum state tomography},\ }\href {https://doi.org/https://doi.org/10.1038/ncomms1147} {\bibfield  {journal} {\bibinfo  {journal} {Nat. Commun.}\ }\textbf {\bibinfo {volume} {1}},\ \bibinfo {pages} {149} (\bibinfo {year} {2010})}\BibitemShut {NoStop}%
\bibitem [{\citenamefont {D'Ariano}\ and\ \citenamefont {Lo~Presti}(2001)}]{ArianoPRL2001}%
  \BibitemOpen
  \bibfield  {author} {\bibinfo {author} {\bibfnamefont {G.~M.}\ \bibnamefont {D'Ariano}}\ and\ \bibinfo {author} {\bibfnamefont {P.}~\bibnamefont {Lo~Presti}},\ }\bibfield  {title} {\bibinfo {title} {Quantum tomography for measuring experimentally the matrix elements of an arbitrary quantum operation},\ }\href {https://doi.org/10.1103/PhysRevLett.86.4195} {\bibfield  {journal} {\bibinfo  {journal} {Phys. Rev. Lett.}\ }\textbf {\bibinfo {volume} {86}},\ \bibinfo {pages} {4195} (\bibinfo {year} {2001})}\BibitemShut {NoStop}%
\bibitem [{\citenamefont {Knill}\ \emph {et~al.}(2008)\citenamefont {Knill}, \citenamefont {Leibfried}, \citenamefont {Reichle}, \citenamefont {Britton}, \citenamefont {Blakestad}, \citenamefont {Jost}, \citenamefont {Langer}, \citenamefont {Ozeri}, \citenamefont {Seidelin},\ and\ \citenamefont {Wineland}}]{KnillPRA2008}%
  \BibitemOpen
  \bibfield  {author} {\bibinfo {author} {\bibfnamefont {E.}~\bibnamefont {Knill}}, \bibinfo {author} {\bibfnamefont {D.}~\bibnamefont {Leibfried}}, \bibinfo {author} {\bibfnamefont {R.}~\bibnamefont {Reichle}}, \bibinfo {author} {\bibfnamefont {J.}~\bibnamefont {Britton}}, \bibinfo {author} {\bibfnamefont {R.~B.}\ \bibnamefont {Blakestad}}, \bibinfo {author} {\bibfnamefont {J.~D.}\ \bibnamefont {Jost}}, \bibinfo {author} {\bibfnamefont {C.}~\bibnamefont {Langer}}, \bibinfo {author} {\bibfnamefont {R.}~\bibnamefont {Ozeri}}, \bibinfo {author} {\bibfnamefont {S.}~\bibnamefont {Seidelin}},\ and\ \bibinfo {author} {\bibfnamefont {D.~J.}\ \bibnamefont {Wineland}},\ }\bibfield  {title} {\bibinfo {title} {Randomized benchmarking of quantum gates},\ }\href {https://doi.org/10.1103/PhysRevA.77.012307} {\bibfield  {journal} {\bibinfo  {journal} {Phys. Rev. A}\ }\textbf {\bibinfo {volume} {77}},\ \bibinfo {pages} {012307} (\bibinfo {year} {2008})}\BibitemShut {NoStop}%
\bibitem [{\citenamefont {Helsen}\ \emph {et~al.}(2022)\citenamefont {Helsen}, \citenamefont {Roth}, \citenamefont {Onorati}, \citenamefont {Werner},\ and\ \citenamefont {Eisert}}]{EisertPRXQ2022}%
  \BibitemOpen
  \bibfield  {author} {\bibinfo {author} {\bibfnamefont {J.}~\bibnamefont {Helsen}}, \bibinfo {author} {\bibfnamefont {I.}~\bibnamefont {Roth}}, \bibinfo {author} {\bibfnamefont {E.}~\bibnamefont {Onorati}}, \bibinfo {author} {\bibfnamefont {A.}~\bibnamefont {Werner}},\ and\ \bibinfo {author} {\bibfnamefont {J.}~\bibnamefont {Eisert}},\ }\bibfield  {title} {\bibinfo {title} {General framework for randomized benchmarking},\ }\href {https://doi.org/10.1103/PRXQuantum.3.020357} {\bibfield  {journal} {\bibinfo  {journal} {PRX Quantum}\ }\textbf {\bibinfo {volume} {3}},\ \bibinfo {pages} {020357} (\bibinfo {year} {2022})}\BibitemShut {NoStop}%
\bibitem [{\citenamefont {Hines}\ \emph {et~al.}(2023)\citenamefont {Hines}, \citenamefont {Lu}, \citenamefont {Naik}, \citenamefont {Hashim}, \citenamefont {Ville}, \citenamefont {Mitchell}, \citenamefont {Kriekebaum}, \citenamefont {Santiago}, \citenamefont {Seritan}, \citenamefont {Nielsen}, \citenamefont {Blume-Kohout}, \citenamefont {Young}, \citenamefont {Siddiqi}, \citenamefont {Whaley},\ and\ \citenamefont {Proctor}}]{HinesPRX2023}%
  \BibitemOpen
  \bibfield  {author} {\bibinfo {author} {\bibfnamefont {J.}~\bibnamefont {Hines}}, \bibinfo {author} {\bibfnamefont {M.}~\bibnamefont {Lu}}, \bibinfo {author} {\bibfnamefont {R.~K.}\ \bibnamefont {Naik}}, \bibinfo {author} {\bibfnamefont {A.}~\bibnamefont {Hashim}}, \bibinfo {author} {\bibfnamefont {J.-L.}\ \bibnamefont {Ville}}, \bibinfo {author} {\bibfnamefont {B.}~\bibnamefont {Mitchell}}, \bibinfo {author} {\bibfnamefont {J.~M.}\ \bibnamefont {Kriekebaum}}, \bibinfo {author} {\bibfnamefont {D.~I.}\ \bibnamefont {Santiago}}, \bibinfo {author} {\bibfnamefont {S.}~\bibnamefont {Seritan}}, \bibinfo {author} {\bibfnamefont {E.}~\bibnamefont {Nielsen}}, \bibinfo {author} {\bibfnamefont {R.}~\bibnamefont {Blume-Kohout}}, \bibinfo {author} {\bibfnamefont {K.}~\bibnamefont {Young}}, \bibinfo {author} {\bibfnamefont {I.}~\bibnamefont {Siddiqi}}, \bibinfo {author} {\bibfnamefont {B.}~\bibnamefont {Whaley}},\ and\ \bibinfo {author} {\bibfnamefont {T.}~\bibnamefont {Proctor}},\ }\bibfield  {title} {\bibinfo {title}
  {Demonstrating scalable randomized benchmarking of universal gate sets},\ }\href {https://doi.org/10.1103/PhysRevX.13.041030} {\bibfield  {journal} {\bibinfo  {journal} {Phys. Rev. X}\ }\textbf {\bibinfo {volume} {13}},\ \bibinfo {pages} {041030} (\bibinfo {year} {2023})}\BibitemShut {NoStop}%
\bibitem [{\citenamefont {Wang}\ \emph {et~al.}(2024)\citenamefont {Wang}, \citenamefont {Liu}, \citenamefont {Liu}, \citenamefont {Gu}, \citenamefont {Baker}, \citenamefont {Chong},\ and\ \citenamefont {Han}}]{WangDGR2024arXiv}%
  \BibitemOpen
  \bibfield  {author} {\bibinfo {author} {\bibfnamefont {H.}~\bibnamefont {Wang}}, \bibinfo {author} {\bibfnamefont {P.}~\bibnamefont {Liu}}, \bibinfo {author} {\bibfnamefont {Y.}~\bibnamefont {Liu}}, \bibinfo {author} {\bibfnamefont {J.}~\bibnamefont {Gu}}, \bibinfo {author} {\bibfnamefont {J.}~\bibnamefont {Baker}}, \bibinfo {author} {\bibfnamefont {F.~T.}\ \bibnamefont {Chong}},\ and\ \bibinfo {author} {\bibfnamefont {S.}~\bibnamefont {Han}},\ }\href {https://arxiv.org/abs/2311.16214} {\bibinfo {title} {Dgr: Tackling drifted and correlated noise in quantum error correction via decoding graph re-weighting}} (\bibinfo {year} {2024}),\ \Eprint {https://arxiv.org/abs/2311.16214} {arXiv:2311.16214 [quant-ph]} \BibitemShut {NoStop}%
\bibitem [{\citenamefont {Chen}\ \emph {et~al.}(2022)\citenamefont {Chen}, \citenamefont {Yoder}, \citenamefont {Kim}, \citenamefont {Sundaresan}, \citenamefont {Srinivasan}, \citenamefont {Li}, \citenamefont {C\'orcoles}, \citenamefont {Cross},\ and\ \citenamefont {Takita}}]{ChenPRL2022}%
  \BibitemOpen
  \bibfield  {author} {\bibinfo {author} {\bibfnamefont {E.~H.}\ \bibnamefont {Chen}}, \bibinfo {author} {\bibfnamefont {T.~J.}\ \bibnamefont {Yoder}}, \bibinfo {author} {\bibfnamefont {Y.}~\bibnamefont {Kim}}, \bibinfo {author} {\bibfnamefont {N.}~\bibnamefont {Sundaresan}}, \bibinfo {author} {\bibfnamefont {S.}~\bibnamefont {Srinivasan}}, \bibinfo {author} {\bibfnamefont {M.}~\bibnamefont {Li}}, \bibinfo {author} {\bibfnamefont {A.~D.}\ \bibnamefont {C\'orcoles}}, \bibinfo {author} {\bibfnamefont {A.~W.}\ \bibnamefont {Cross}},\ and\ \bibinfo {author} {\bibfnamefont {M.}~\bibnamefont {Takita}},\ }\bibfield  {title} {\bibinfo {title} {Calibrated decoders for experimental quantum error correction},\ }\href {https://doi.org/10.1103/PhysRevLett.128.110504} {\bibfield  {journal} {\bibinfo  {journal} {Phys. Rev. Lett.}\ }\textbf {\bibinfo {volume} {128}},\ \bibinfo {pages} {110504} (\bibinfo {year} {2022})}\BibitemShut {NoStop}%
\bibitem [{\citenamefont {Remm}\ \emph {et~al.}(2025)\citenamefont {Remm}, \citenamefont {Lacroix}, \citenamefont {Bödeker}, \citenamefont {Genois}, \citenamefont {Hellings}, \citenamefont {Swiadek}, \citenamefont {Norris}, \citenamefont {Eichler}, \citenamefont {Blais}, \citenamefont {Müller}, \citenamefont {Krinner},\ and\ \citenamefont {Wallraff}}]{WallraffArXiv2025}%
  \BibitemOpen
  \bibfield  {author} {\bibinfo {author} {\bibfnamefont {A.}~\bibnamefont {Remm}}, \bibinfo {author} {\bibfnamefont {N.}~\bibnamefont {Lacroix}}, \bibinfo {author} {\bibfnamefont {L.}~\bibnamefont {Bödeker}}, \bibinfo {author} {\bibfnamefont {E.}~\bibnamefont {Genois}}, \bibinfo {author} {\bibfnamefont {C.}~\bibnamefont {Hellings}}, \bibinfo {author} {\bibfnamefont {F.}~\bibnamefont {Swiadek}}, \bibinfo {author} {\bibfnamefont {G.~J.}\ \bibnamefont {Norris}}, \bibinfo {author} {\bibfnamefont {C.}~\bibnamefont {Eichler}}, \bibinfo {author} {\bibfnamefont {A.}~\bibnamefont {Blais}}, \bibinfo {author} {\bibfnamefont {M.}~\bibnamefont {Müller}}, \bibinfo {author} {\bibfnamefont {S.}~\bibnamefont {Krinner}},\ and\ \bibinfo {author} {\bibfnamefont {A.}~\bibnamefont {Wallraff}},\ }\href {https://arxiv.org/abs/2502.17722} {\bibinfo {title} {Experimentally informed decoding of stabilizer codes based on syndrome correlations}} (\bibinfo {year} {2025}),\ \Eprint {https://arxiv.org/abs/2502.17722} {arXiv:2502.17722
  [quant-ph]} \BibitemShut {NoStop}%
\bibitem [{\citenamefont {Tuckett}\ \emph {et~al.}(2019)\citenamefont {Tuckett}, \citenamefont {Darmawan}, \citenamefont {Chubb}, \citenamefont {Bravyi}, \citenamefont {Bartlett},\ and\ \citenamefont {Flammia}}]{FlamiaPRX2019}%
  \BibitemOpen
  \bibfield  {author} {\bibinfo {author} {\bibfnamefont {D.~K.}\ \bibnamefont {Tuckett}}, \bibinfo {author} {\bibfnamefont {A.~S.}\ \bibnamefont {Darmawan}}, \bibinfo {author} {\bibfnamefont {C.~T.}\ \bibnamefont {Chubb}}, \bibinfo {author} {\bibfnamefont {S.}~\bibnamefont {Bravyi}}, \bibinfo {author} {\bibfnamefont {S.~D.}\ \bibnamefont {Bartlett}},\ and\ \bibinfo {author} {\bibfnamefont {S.~T.}\ \bibnamefont {Flammia}},\ }\bibfield  {title} {\bibinfo {title} {Tailoring surface codes for highly biased noise},\ }\href {https://doi.org/10.1103/PhysRevX.9.041031} {\bibfield  {journal} {\bibinfo  {journal} {Phys. Rev. X}\ }\textbf {\bibinfo {volume} {9}},\ \bibinfo {pages} {041031} (\bibinfo {year} {2019})}\BibitemShut {NoStop}%
\bibitem [{\citenamefont {Tuckett}\ \emph {et~al.}(2020)\citenamefont {Tuckett}, \citenamefont {Bartlett}, \citenamefont {Flammia},\ and\ \citenamefont {Brown}}]{TuckettPRL2020}%
  \BibitemOpen
  \bibfield  {author} {\bibinfo {author} {\bibfnamefont {D.~K.}\ \bibnamefont {Tuckett}}, \bibinfo {author} {\bibfnamefont {S.~D.}\ \bibnamefont {Bartlett}}, \bibinfo {author} {\bibfnamefont {S.~T.}\ \bibnamefont {Flammia}},\ and\ \bibinfo {author} {\bibfnamefont {B.~J.}\ \bibnamefont {Brown}},\ }\bibfield  {title} {\bibinfo {title} {Fault-tolerant thresholds for the surface code in excess of 5\% under biased noise},\ }\href {https://doi.org/10.1103/PhysRevLett.124.130501} {\bibfield  {journal} {\bibinfo  {journal} {Phys. Rev. Lett.}\ }\textbf {\bibinfo {volume} {124}},\ \bibinfo {pages} {130501} (\bibinfo {year} {2020})}\BibitemShut {NoStop}%
\bibitem [{\citenamefont {Huang}\ \emph {et~al.}(2023)\citenamefont {Huang}, \citenamefont {Pesah}, \citenamefont {Chubb}, \citenamefont {Vasmer},\ and\ \citenamefont {Dua}}]{ArpitPRXQ2023}%
  \BibitemOpen
  \bibfield  {author} {\bibinfo {author} {\bibfnamefont {E.}~\bibnamefont {Huang}}, \bibinfo {author} {\bibfnamefont {A.}~\bibnamefont {Pesah}}, \bibinfo {author} {\bibfnamefont {C.~T.}\ \bibnamefont {Chubb}}, \bibinfo {author} {\bibfnamefont {M.}~\bibnamefont {Vasmer}},\ and\ \bibinfo {author} {\bibfnamefont {A.}~\bibnamefont {Dua}},\ }\bibfield  {title} {\bibinfo {title} {Tailoring three-dimensional topological codes for biased noise},\ }\href {https://doi.org/10.1103/PRXQuantum.4.030338} {\bibfield  {journal} {\bibinfo  {journal} {PRX Quantum}\ }\textbf {\bibinfo {volume} {4}},\ \bibinfo {pages} {030338} (\bibinfo {year} {2023})}\BibitemShut {NoStop}%
\bibitem [{\citenamefont {Campos}\ and\ \citenamefont {Brown}(2024)}]{Campos2024arXiv}%
  \BibitemOpen
  \bibfield  {author} {\bibinfo {author} {\bibfnamefont {J.~A.}\ \bibnamefont {Campos}}\ and\ \bibinfo {author} {\bibfnamefont {K.~R.}\ \bibnamefont {Brown}},\ }\href {https://arxiv.org/abs/2412.03808} {\bibinfo {title} {Clifford-deformed compass codes}} (\bibinfo {year} {2024}),\ \Eprint {https://arxiv.org/abs/2412.03808} {arXiv:2412.03808 [quant-ph]} \BibitemShut {NoStop}%
\bibitem [{\citenamefont {Song}\ \emph {et~al.}(2017)\citenamefont {Song}, \citenamefont {Xu}, \citenamefont {Liu}, \citenamefont {Yang}, \citenamefont {Zheng}, \citenamefont {Deng}, \citenamefont {Xie}, \citenamefont {Huang}, \citenamefont {Guo}, \citenamefont {Zhang}, \citenamefont {Zhang}, \citenamefont {Xu}, \citenamefont {Zheng}, \citenamefont {Zhu}, \citenamefont {Wang}, \citenamefont {Chen}, \citenamefont {Lu}, \citenamefont {Han},\ and\ \citenamefont {Pan}}]{PanPRL2017}%
  \BibitemOpen
  \bibfield  {author} {\bibinfo {author} {\bibfnamefont {C.}~\bibnamefont {Song}}, \bibinfo {author} {\bibfnamefont {K.}~\bibnamefont {Xu}}, \bibinfo {author} {\bibfnamefont {W.}~\bibnamefont {Liu}}, \bibinfo {author} {\bibfnamefont {C.-p.}\ \bibnamefont {Yang}}, \bibinfo {author} {\bibfnamefont {S.-B.}\ \bibnamefont {Zheng}}, \bibinfo {author} {\bibfnamefont {H.}~\bibnamefont {Deng}}, \bibinfo {author} {\bibfnamefont {Q.}~\bibnamefont {Xie}}, \bibinfo {author} {\bibfnamefont {K.}~\bibnamefont {Huang}}, \bibinfo {author} {\bibfnamefont {Q.}~\bibnamefont {Guo}}, \bibinfo {author} {\bibfnamefont {L.}~\bibnamefont {Zhang}}, \bibinfo {author} {\bibfnamefont {P.}~\bibnamefont {Zhang}}, \bibinfo {author} {\bibfnamefont {D.}~\bibnamefont {Xu}}, \bibinfo {author} {\bibfnamefont {D.}~\bibnamefont {Zheng}}, \bibinfo {author} {\bibfnamefont {X.}~\bibnamefont {Zhu}}, \bibinfo {author} {\bibfnamefont {H.}~\bibnamefont {Wang}}, \bibinfo {author} {\bibfnamefont {Y.-A.}\ \bibnamefont {Chen}}, \bibinfo {author} {\bibfnamefont
  {C.-Y.}\ \bibnamefont {Lu}}, \bibinfo {author} {\bibfnamefont {S.}~\bibnamefont {Han}},\ and\ \bibinfo {author} {\bibfnamefont {J.-W.}\ \bibnamefont {Pan}},\ }\bibfield  {title} {\bibinfo {title} {10-qubit entanglement and parallel logic operations with a superconducting circuit},\ }\href {https://doi.org/10.1103/PhysRevLett.119.180511} {\bibfield  {journal} {\bibinfo  {journal} {Phys. Rev. Lett.}\ }\textbf {\bibinfo {volume} {119}},\ \bibinfo {pages} {180511} (\bibinfo {year} {2017})}\BibitemShut {NoStop}%
\bibitem [{\citenamefont {Aaronson}(2018)}]{AaronsonArXiv2018}%
  \BibitemOpen
  \bibfield  {author} {\bibinfo {author} {\bibfnamefont {S.}~\bibnamefont {Aaronson}},\ }\href {https://arxiv.org/abs/1711.01053} {\bibinfo {title} {Shadow tomography of quantum states}} (\bibinfo {year} {2018}),\ \Eprint {https://arxiv.org/abs/1711.01053} {arXiv:1711.01053 [quant-ph]} \BibitemShut {NoStop}%
\bibitem [{\citenamefont {Hu}\ \emph {et~al.}(2024)\citenamefont {Hu}, \citenamefont {Wei}, \citenamefont {Liu}, \citenamefont {Che}, \citenamefont {Zhou}, \citenamefont {Xie}, \citenamefont {Qin}, \citenamefont {Hu}, \citenamefont {Yuan}, \citenamefont {Zhou}, \citenamefont {Liu}, \citenamefont {Tan}, \citenamefont {Xin},\ and\ \citenamefont {Yu}}]{DapengPRL2024}%
  \BibitemOpen
  \bibfield  {author} {\bibinfo {author} {\bibfnamefont {C.-K.}\ \bibnamefont {Hu}}, \bibinfo {author} {\bibfnamefont {C.}~\bibnamefont {Wei}}, \bibinfo {author} {\bibfnamefont {C.}~\bibnamefont {Liu}}, \bibinfo {author} {\bibfnamefont {L.}~\bibnamefont {Che}}, \bibinfo {author} {\bibfnamefont {Y.}~\bibnamefont {Zhou}}, \bibinfo {author} {\bibfnamefont {G.}~\bibnamefont {Xie}}, \bibinfo {author} {\bibfnamefont {H.}~\bibnamefont {Qin}}, \bibinfo {author} {\bibfnamefont {G.}~\bibnamefont {Hu}}, \bibinfo {author} {\bibfnamefont {H.}~\bibnamefont {Yuan}}, \bibinfo {author} {\bibfnamefont {R.}~\bibnamefont {Zhou}}, \bibinfo {author} {\bibfnamefont {S.}~\bibnamefont {Liu}}, \bibinfo {author} {\bibfnamefont {D.}~\bibnamefont {Tan}}, \bibinfo {author} {\bibfnamefont {T.}~\bibnamefont {Xin}},\ and\ \bibinfo {author} {\bibfnamefont {D.}~\bibnamefont {Yu}},\ }\bibfield  {title} {\bibinfo {title} {Experimental sample-efficient quantum state tomography via parallel measurements},\ }\href
  {https://doi.org/10.1103/PhysRevLett.133.160801} {\bibfield  {journal} {\bibinfo  {journal} {Phys. Rev. Lett.}\ }\textbf {\bibinfo {volume} {133}},\ \bibinfo {pages} {160801} (\bibinfo {year} {2024})}\BibitemShut {NoStop}%
\bibitem [{\citenamefont {Gross}\ \emph {et~al.}(2010)\citenamefont {Gross}, \citenamefont {Liu}, \citenamefont {Flammia}, \citenamefont {Becker},\ and\ \citenamefont {Eisert}}]{EisertPRL2010}%
  \BibitemOpen
  \bibfield  {author} {\bibinfo {author} {\bibfnamefont {D.}~\bibnamefont {Gross}}, \bibinfo {author} {\bibfnamefont {Y.-K.}\ \bibnamefont {Liu}}, \bibinfo {author} {\bibfnamefont {S.~T.}\ \bibnamefont {Flammia}}, \bibinfo {author} {\bibfnamefont {S.}~\bibnamefont {Becker}},\ and\ \bibinfo {author} {\bibfnamefont {J.}~\bibnamefont {Eisert}},\ }\bibfield  {title} {\bibinfo {title} {Quantum state tomography via compressed sensing},\ }\href {https://doi.org/10.1103/PhysRevLett.105.150401} {\bibfield  {journal} {\bibinfo  {journal} {Phys. Rev. Lett.}\ }\textbf {\bibinfo {volume} {105}},\ \bibinfo {pages} {150401} (\bibinfo {year} {2010})}\BibitemShut {NoStop}%
\bibitem [{\citenamefont {Xin}\ \emph {et~al.}(2017)\citenamefont {Xin}, \citenamefont {Lu}, \citenamefont {Klassen}, \citenamefont {Yu}, \citenamefont {Ji}, \citenamefont {Chen}, \citenamefont {Ma}, \citenamefont {Long}, \citenamefont {Zeng},\ and\ \citenamefont {Laflamme}}]{LaflammePRL2017}%
  \BibitemOpen
  \bibfield  {author} {\bibinfo {author} {\bibfnamefont {T.}~\bibnamefont {Xin}}, \bibinfo {author} {\bibfnamefont {D.}~\bibnamefont {Lu}}, \bibinfo {author} {\bibfnamefont {J.}~\bibnamefont {Klassen}}, \bibinfo {author} {\bibfnamefont {N.}~\bibnamefont {Yu}}, \bibinfo {author} {\bibfnamefont {Z.}~\bibnamefont {Ji}}, \bibinfo {author} {\bibfnamefont {J.}~\bibnamefont {Chen}}, \bibinfo {author} {\bibfnamefont {X.}~\bibnamefont {Ma}}, \bibinfo {author} {\bibfnamefont {G.}~\bibnamefont {Long}}, \bibinfo {author} {\bibfnamefont {B.}~\bibnamefont {Zeng}},\ and\ \bibinfo {author} {\bibfnamefont {R.}~\bibnamefont {Laflamme}},\ }\bibfield  {title} {\bibinfo {title} {Quantum state tomography via reduced density matrices},\ }\href {https://doi.org/10.1103/PhysRevLett.118.020401} {\bibfield  {journal} {\bibinfo  {journal} {Phys. Rev. Lett.}\ }\textbf {\bibinfo {volume} {118}},\ \bibinfo {pages} {020401} (\bibinfo {year} {2017})}\BibitemShut {NoStop}%
\bibitem [{\citenamefont {Flammia}(2022)}]{FlamiaACES2022}%
  \BibitemOpen
  \bibfield  {author} {\bibinfo {author} {\bibfnamefont {S.~T.}\ \bibnamefont {Flammia}},\ }\bibfield  {title} {\bibinfo {title} {{Averaged Circuit Eigenvalue Sampling}},\ }in\ \href {https://doi.org/10.4230/LIPIcs.TQC.2022.4} {\emph {\bibinfo {booktitle} {17th Conference on the Theory of Quantum Computation, Communication and Cryptography (TQC 2022)}}},\ \bibinfo {series} {Leibniz International Proceedings in Informatics (LIPIcs)}, Vol.\ \bibinfo {volume} {232},\ \bibinfo {editor} {edited by\ \bibinfo {editor} {\bibfnamefont {F.}~\bibnamefont {Le~Gall}}\ and\ \bibinfo {editor} {\bibfnamefont {T.}~\bibnamefont {Morimae}}}\ (\bibinfo  {publisher} {Schloss Dagstuhl -- Leibniz-Zentrum f{\"u}r Informatik},\ \bibinfo {address} {Dagstuhl, Germany},\ \bibinfo {year} {2022})\ pp.\ \bibinfo {pages} {4:1--4:10}\BibitemShut {NoStop}%
\bibitem [{\citenamefont {Hockings}\ \emph {et~al.}(2025{\natexlab{a}})\citenamefont {Hockings}, \citenamefont {Doherty},\ and\ \citenamefont {Harper}}]{HarperPRXQ2025}%
  \BibitemOpen
  \bibfield  {author} {\bibinfo {author} {\bibfnamefont {E.~T.}\ \bibnamefont {Hockings}}, \bibinfo {author} {\bibfnamefont {A.~C.}\ \bibnamefont {Doherty}},\ and\ \bibinfo {author} {\bibfnamefont {R.}~\bibnamefont {Harper}},\ }\bibfield  {title} {\bibinfo {title} {Scalable noise characterization of syndrome-extraction circuits with averaged circuit eigenvalue sampling},\ }\href {https://doi.org/10.1103/PRXQuantum.6.010334} {\bibfield  {journal} {\bibinfo  {journal} {PRX Quantum}\ }\textbf {\bibinfo {volume} {6}},\ \bibinfo {pages} {010334} (\bibinfo {year} {2025}{\natexlab{a}})}\BibitemShut {NoStop}%
\bibitem [{\citenamefont {Hockings}\ \emph {et~al.}(2025{\natexlab{b}})\citenamefont {Hockings}, \citenamefont {Doherty},\ and\ \citenamefont {Harper}}]{Harper2025arXiv}%
  \BibitemOpen
  \bibfield  {author} {\bibinfo {author} {\bibfnamefont {E.~T.}\ \bibnamefont {Hockings}}, \bibinfo {author} {\bibfnamefont {A.~C.}\ \bibnamefont {Doherty}},\ and\ \bibinfo {author} {\bibfnamefont {R.}~\bibnamefont {Harper}},\ }\href {https://arxiv.org/abs/2502.21044} {\bibinfo {title} {Improving error suppression with noise-aware decoding}} (\bibinfo {year} {2025}{\natexlab{b}}),\ \Eprint {https://arxiv.org/abs/2502.21044} {arXiv:2502.21044 [quant-ph]} \BibitemShut {NoStop}%
\bibitem [{\citenamefont {Harper}\ and\ \citenamefont {Flammia}(2023)}]{FlamiaPRXQ2023}%
  \BibitemOpen
  \bibfield  {author} {\bibinfo {author} {\bibfnamefont {R.}~\bibnamefont {Harper}}\ and\ \bibinfo {author} {\bibfnamefont {S.~T.}\ \bibnamefont {Flammia}},\ }\bibfield  {title} {\bibinfo {title} {Learning correlated noise in a 39-qubit quantum processor},\ }\href {https://doi.org/10.1103/PRXQuantum.4.040311} {\bibfield  {journal} {\bibinfo  {journal} {PRX Quantum}\ }\textbf {\bibinfo {volume} {4}},\ \bibinfo {pages} {040311} (\bibinfo {year} {2023})}\BibitemShut {NoStop}%
\bibitem [{\citenamefont {Chen}\ \emph {et~al.}(2023)\citenamefont {Chen}, \citenamefont {Liu}, \citenamefont {Otten}, \citenamefont {Seif}, \citenamefont {Fefferman},\ and\ \citenamefont {Jiang}}]{JiangNatCommun2023}%
  \BibitemOpen
  \bibfield  {author} {\bibinfo {author} {\bibfnamefont {S.}~\bibnamefont {Chen}}, \bibinfo {author} {\bibfnamefont {Y.}~\bibnamefont {Liu}}, \bibinfo {author} {\bibfnamefont {M.}~\bibnamefont {Otten}}, \bibinfo {author} {\bibfnamefont {A.}~\bibnamefont {Seif}}, \bibinfo {author} {\bibfnamefont {B.}~\bibnamefont {Fefferman}},\ and\ \bibinfo {author} {\bibfnamefont {L.}~\bibnamefont {Jiang}},\ }\bibfield  {title} {\bibinfo {title} {The learnability of pauli noise},\ }\href {https://doi.org/https://doi.org/10.1038/s41467-022-35759-4} {\bibfield  {journal} {\bibinfo  {journal} {Nat Commun.}\ }\textbf {\bibinfo {volume} {14}},\ \bibinfo {pages} {52} (\bibinfo {year} {2023})}\BibitemShut {NoStop}%
\bibitem [{\citenamefont {Chen}\ \emph {et~al.}(2024)\citenamefont {Chen}, \citenamefont {Zhang}, \citenamefont {Jiang},\ and\ \citenamefont {Flammia}}]{Chen2024ArXiv}%
  \BibitemOpen
  \bibfield  {author} {\bibinfo {author} {\bibfnamefont {S.}~\bibnamefont {Chen}}, \bibinfo {author} {\bibfnamefont {Z.}~\bibnamefont {Zhang}}, \bibinfo {author} {\bibfnamefont {L.}~\bibnamefont {Jiang}},\ and\ \bibinfo {author} {\bibfnamefont {S.~T.}\ \bibnamefont {Flammia}},\ }\href {https://arxiv.org/abs/2410.03906} {\bibinfo {title} {Efficient self-consistent learning of gate set pauli noise}} (\bibinfo {year} {2024}),\ \Eprint {https://arxiv.org/abs/2410.03906} {arXiv:2410.03906 [quant-ph]} \BibitemShut {NoStop}%
\bibitem [{\citenamefont {Spitz}\ \emph {et~al.}(2018)\citenamefont {Spitz}, \citenamefont {Tarasinski}, \citenamefont {Beenakker},\ and\ \citenamefont {O'Brien}}]{Spitz2018}%
  \BibitemOpen
  \bibfield  {author} {\bibinfo {author} {\bibfnamefont {S.~T.}\ \bibnamefont {Spitz}}, \bibinfo {author} {\bibfnamefont {B.}~\bibnamefont {Tarasinski}}, \bibinfo {author} {\bibfnamefont {C.~W.~J.}\ \bibnamefont {Beenakker}},\ and\ \bibinfo {author} {\bibfnamefont {T.~E.}\ \bibnamefont {O'Brien}},\ }\bibfield  {title} {\bibinfo {title} {Adaptive weight estimator for quantum error correction in a time-dependent environment},\ }\href {https://doi.org/https://doi.org/10.1002/qute.201800012} {\bibfield  {journal} {\bibinfo  {journal} {Advanced Quantum Technologies}\ }\textbf {\bibinfo {volume} {1}},\ \bibinfo {pages} {1800012} (\bibinfo {year} {2018})}\BibitemShut {NoStop}%
\bibitem [{\citenamefont {Wagner}\ \emph {et~al.}(2021)\citenamefont {Wagner}, \citenamefont {Kampermann}, \citenamefont {Bru\ss{}},\ and\ \citenamefont {Kliesch}}]{WagnerPRR2021}%
  \BibitemOpen
  \bibfield  {author} {\bibinfo {author} {\bibfnamefont {T.}~\bibnamefont {Wagner}}, \bibinfo {author} {\bibfnamefont {H.}~\bibnamefont {Kampermann}}, \bibinfo {author} {\bibfnamefont {D.}~\bibnamefont {Bru\ss{}}},\ and\ \bibinfo {author} {\bibfnamefont {M.}~\bibnamefont {Kliesch}},\ }\bibfield  {title} {\bibinfo {title} {Optimal noise estimation from syndrome statistics of quantum codes},\ }\href {https://doi.org/10.1103/PhysRevResearch.3.013292} {\bibfield  {journal} {\bibinfo  {journal} {Phys. Rev. Res.}\ }\textbf {\bibinfo {volume} {3}},\ \bibinfo {pages} {013292} (\bibinfo {year} {2021})}\BibitemShut {NoStop}%
\bibitem [{\citenamefont {AI}(2021)}]{GoogleNature2021}%
  \BibitemOpen
  \bibfield  {author} {\bibinfo {author} {\bibfnamefont {G.~Q.}\ \bibnamefont {AI}},\ }\bibfield  {title} {\bibinfo {title} {Exponential suppression of bit or phase errors with cyclic error correction},\ }\href {https://doi.org/https://doi.org/10.1038/s41586-021-03588-y} {\bibfield  {journal} {\bibinfo  {journal} {Nature}\ }\textbf {\bibinfo {volume} {595}},\ \bibinfo {pages} {383–387} (\bibinfo {year} {2021})}\BibitemShut {NoStop}%
\bibitem [{\citenamefont {AI}(2023)}]{GoogleNature2024}%
  \BibitemOpen
  \bibfield  {author} {\bibinfo {author} {\bibfnamefont {G.~Q.}\ \bibnamefont {AI}},\ }\bibfield  {title} {\bibinfo {title} {Suppressing quantum errors by scaling a surface code logical qubit},\ }\href {https://doi.org/https://doi.org/10.1038/s41586-022-05434-1} {\bibfield  {journal} {\bibinfo  {journal} {Nature}\ }\textbf {\bibinfo {volume} {614}},\ \bibinfo {pages} {676–681} (\bibinfo {year} {2023})}\BibitemShut {NoStop}%
\bibitem [{\citenamefont {Gidney}(2021)}]{Gidney2021Stim}%
  \BibitemOpen
  \bibfield  {author} {\bibinfo {author} {\bibfnamefont {C.}~\bibnamefont {Gidney}},\ }\bibfield  {title} {\bibinfo {title} {Stim: a fast stabilizer circuit simulator},\ }\href {https://doi.org/10.22331/q-2021-07-06-497} {\bibfield  {journal} {\bibinfo  {journal} {{Quantum}}\ }\textbf {\bibinfo {volume} {5}},\ \bibinfo {pages} {497} (\bibinfo {year} {2021})}\BibitemShut {NoStop}%
\bibitem [{\citenamefont {Higgott}\ and\ \citenamefont {Gidney}(2025)}]{HiggotQuantum2025}%
  \BibitemOpen
  \bibfield  {author} {\bibinfo {author} {\bibfnamefont {O.}~\bibnamefont {Higgott}}\ and\ \bibinfo {author} {\bibfnamefont {C.}~\bibnamefont {Gidney}},\ }\bibfield  {title} {\bibinfo {title} {Sparse {B}lossom: correcting a million errors per core second with minimum-weight matching},\ }\href {https://doi.org/10.22331/q-2025-01-20-1600} {\bibfield  {journal} {\bibinfo  {journal} {{Quantum}}\ }\textbf {\bibinfo {volume} {9}},\ \bibinfo {pages} {1600} (\bibinfo {year} {2025})}\BibitemShut {NoStop}%
\bibitem [{\citenamefont {Derks}\ \emph {et~al.}(2024)\citenamefont {Derks}, \citenamefont {Townsend-Teague}, \citenamefont {Burchards},\ and\ \citenamefont {Eisert}}]{EisertarXiv2024}%
  \BibitemOpen
  \bibfield  {author} {\bibinfo {author} {\bibfnamefont {P.-J. H.~S.}\ \bibnamefont {Derks}}, \bibinfo {author} {\bibfnamefont {A.}~\bibnamefont {Townsend-Teague}}, \bibinfo {author} {\bibfnamefont {A.~G.}\ \bibnamefont {Burchards}},\ and\ \bibinfo {author} {\bibfnamefont {J.}~\bibnamefont {Eisert}},\ }\href {https://arxiv.org/abs/2407.13826} {\bibinfo {title} {Designing fault-tolerant circuits using detector error models}} (\bibinfo {year} {2024}),\ \Eprint {https://arxiv.org/abs/2407.13826} {arXiv:2407.13826 [quant-ph]} \BibitemShut {NoStop}%
\bibitem [{\citenamefont {Edmonds}(1965)}]{Edmonds_1965}%
  \BibitemOpen
  \bibfield  {author} {\bibinfo {author} {\bibfnamefont {J.}~\bibnamefont {Edmonds}},\ }\bibfield  {title} {\bibinfo {title} {Paths, trees, and flowers},\ }\href {https://doi.org/10.4153/CJM-1965-045-4} {\bibfield  {journal} {\bibinfo  {journal} {Canadian Journal of Mathematics}\ }\textbf {\bibinfo {volume} {17}},\ \bibinfo {pages} {449–467} (\bibinfo {year} {1965})}\BibitemShut {NoStop}%
\bibitem [{\citenamefont {Higgott}(2022)}]{HiggotACM2022}%
  \BibitemOpen
  \bibfield  {author} {\bibinfo {author} {\bibfnamefont {O.}~\bibnamefont {Higgott}},\ }\bibfield  {title} {\bibinfo {title} {Pymatching: A python package for decoding quantum codes with minimum-weight perfect matching},\ }\bibfield  {journal} {\bibinfo  {journal} {ACM T QUANTUM COMPUT}\ }\textbf {\bibinfo {volume} {3}},\ \href {https://doi.org/10.1145/3505637} {10.1145/3505637} (\bibinfo {year} {2022})\BibitemShut {NoStop}%
\bibitem [{\citenamefont {Nielsen}(2002)}]{NIELSENPhysicsLettersA2002}%
  \BibitemOpen
  \bibfield  {author} {\bibinfo {author} {\bibfnamefont {M.~A.}\ \bibnamefont {Nielsen}},\ }\bibfield  {title} {\bibinfo {title} {A simple formula for the average gate fidelity of a quantum dynamical operation},\ }\href {https://doi.org/https://doi.org/10.1016/S0375-9601(02)01272-0} {\bibfield  {journal} {\bibinfo  {journal} {Physics Letters A}\ }\textbf {\bibinfo {volume} {303}},\ \bibinfo {pages} {249} (\bibinfo {year} {2002})}\BibitemShut {NoStop}%
\bibitem [{\citenamefont {Greenbaum}\ and\ \citenamefont {Dutton}(2017)}]{DuttonQST2017}%
  \BibitemOpen
  \bibfield  {author} {\bibinfo {author} {\bibfnamefont {D.}~\bibnamefont {Greenbaum}}\ and\ \bibinfo {author} {\bibfnamefont {Z.}~\bibnamefont {Dutton}},\ }\bibfield  {title} {\bibinfo {title} {Modeling coherent errors in quantum error correction},\ }\href {https://doi.org/10.1088/2058-9565/aa9a06} {\bibfield  {journal} {\bibinfo  {journal} {Quantum Sci. Technol.}\ }\textbf {\bibinfo {volume} {3}},\ \bibinfo {pages} {015007} (\bibinfo {year} {2017})}\BibitemShut {NoStop}%
\bibitem [{\citenamefont {Paz-Silva}\ and\ \citenamefont {Lidar}(2013)}]{LidarSciRep2013}%
  \BibitemOpen
  \bibfield  {author} {\bibinfo {author} {\bibfnamefont {G.~A.}\ \bibnamefont {Paz-Silva}}\ and\ \bibinfo {author} {\bibfnamefont {D.~A.}\ \bibnamefont {Lidar}},\ }\bibfield  {title} {\bibinfo {title} {Optimally combining dynamical decoupling and quantum error correction},\ }\href {https://doi.org/https://doi.org/10.1038/srep01530} {\bibfield  {journal} {\bibinfo  {journal} {Sci. Rep.}\ }\textbf {\bibinfo {volume} {3}},\ \bibinfo {pages} {1530} (\bibinfo {year} {2013})}\BibitemShut {NoStop}%
\bibitem [{\citenamefont {Khodjasteh}\ and\ \citenamefont {Lidar}(2005)}]{LidarPRL2005}%
  \BibitemOpen
  \bibfield  {author} {\bibinfo {author} {\bibfnamefont {K.}~\bibnamefont {Khodjasteh}}\ and\ \bibinfo {author} {\bibfnamefont {D.~A.}\ \bibnamefont {Lidar}},\ }\bibfield  {title} {\bibinfo {title} {Fault-tolerant quantum dynamical decoupling},\ }\href {https://doi.org/10.1103/PhysRevLett.95.180501} {\bibfield  {journal} {\bibinfo  {journal} {Phys. Rev. Lett.}\ }\textbf {\bibinfo {volume} {95}},\ \bibinfo {pages} {180501} (\bibinfo {year} {2005})}\BibitemShut {NoStop}%
\bibitem [{\citenamefont {Jiang}\ and\ \citenamefont {Imambekov}(2011)}]{JiangPRA2011}%
  \BibitemOpen
  \bibfield  {author} {\bibinfo {author} {\bibfnamefont {L.}~\bibnamefont {Jiang}}\ and\ \bibinfo {author} {\bibfnamefont {A.}~\bibnamefont {Imambekov}},\ }\bibfield  {title} {\bibinfo {title} {Universal dynamical decoupling of multiqubit states from environment},\ }\href {https://doi.org/10.1103/PhysRevA.84.060302} {\bibfield  {journal} {\bibinfo  {journal} {Phys. Rev. A}\ }\textbf {\bibinfo {volume} {84}},\ \bibinfo {pages} {060302} (\bibinfo {year} {2011})}\BibitemShut {NoStop}%
\bibitem [{\citenamefont {Zhou}\ \emph {et~al.}(2025)\citenamefont {Zhou}, \citenamefont {Ji},\ and\ \citenamefont {Ding}}]{zhou2025arXiv}%
  \BibitemOpen
  \bibfield  {author} {\bibinfo {author} {\bibfnamefont {Z.}~\bibnamefont {Zhou}}, \bibinfo {author} {\bibfnamefont {A.}~\bibnamefont {Ji}},\ and\ \bibinfo {author} {\bibfnamefont {Y.}~\bibnamefont {Ding}},\ }\href {https://arxiv.org/abs/2503.04642} {\bibinfo {title} {Characterization and mitigation of crosstalk in quantum error correction}} (\bibinfo {year} {2025}),\ \Eprint {https://arxiv.org/abs/2503.04642} {arXiv:2503.04642 [quant-ph]} \BibitemShut {NoStop}%
\bibitem [{\citenamefont {Takou}(2025)}]{Takou2025github}%
  \BibitemOpen
  \bibfield  {author} {\bibinfo {author} {\bibfnamefont {E.}~\bibnamefont {Takou}},\ }\href@noop {} {\bibinfo {title} {``code to simulate noise estimation of dems"}},\ \bibinfo {howpublished} {\url{https://github.com/eva-takou/Noise_estimation_of_DEMs}} (\bibinfo {year} {2025})\BibitemShut {NoStop}%
\bibitem [{\citenamefont {Steane}(1997)}]{SteanePRL1997}%
  \BibitemOpen
  \bibfield  {author} {\bibinfo {author} {\bibfnamefont {A.~M.}\ \bibnamefont {Steane}},\ }\bibfield  {title} {\bibinfo {title} {Active stabilization, quantum computation, and quantum state synthesis},\ }\href {https://doi.org/10.1103/PhysRevLett.78.2252} {\bibfield  {journal} {\bibinfo  {journal} {Phys. Rev. Lett.}\ }\textbf {\bibinfo {volume} {78}},\ \bibinfo {pages} {2252} (\bibinfo {year} {1997})}\BibitemShut {NoStop}%
\bibitem [{\citenamefont {Steane}(2004)}]{Steane2004arXiv}%
  \BibitemOpen
  \bibfield  {author} {\bibinfo {author} {\bibfnamefont {A.~M.}\ \bibnamefont {Steane}},\ }\href {https://arxiv.org/abs/quant-ph/0202036} {\bibinfo {title} {Fast fault-tolerant filtering of quantum codewords}} (\bibinfo {year} {2004}),\ \Eprint {https://arxiv.org/abs/quant-ph/0202036} {arXiv:quant-ph/0202036 [quant-ph]} \BibitemShut {NoStop}%
\bibitem [{\citenamefont {Huang}\ and\ \citenamefont {Brown}(2021)}]{HuangPRL2021}%
  \BibitemOpen
  \bibfield  {author} {\bibinfo {author} {\bibfnamefont {S.}~\bibnamefont {Huang}}\ and\ \bibinfo {author} {\bibfnamefont {K.~R.}\ \bibnamefont {Brown}},\ }\bibfield  {title} {\bibinfo {title} {Between shor and steane: A unifying construction for measuring error syndromes},\ }\href {https://doi.org/10.1103/PhysRevLett.127.090505} {\bibfield  {journal} {\bibinfo  {journal} {Phys. Rev. Lett.}\ }\textbf {\bibinfo {volume} {127}},\ \bibinfo {pages} {090505} (\bibinfo {year} {2021})}\BibitemShut {NoStop}%
\bibitem [{\citenamefont {Postler}\ \emph {et~al.}(2024)\citenamefont {Postler}, \citenamefont {Butt}, \citenamefont {Pogorelov}, \citenamefont {Marciniak}, \citenamefont {Heu\ss{}en}, \citenamefont {Blatt}, \citenamefont {Schindler}, \citenamefont {Rispler}, \citenamefont {M\"uller},\ and\ \citenamefont {Monz}}]{PostlerPRXQuantum2024}%
  \BibitemOpen
  \bibfield  {author} {\bibinfo {author} {\bibfnamefont {L.}~\bibnamefont {Postler}}, \bibinfo {author} {\bibfnamefont {F.}~\bibnamefont {Butt}}, \bibinfo {author} {\bibfnamefont {I.}~\bibnamefont {Pogorelov}}, \bibinfo {author} {\bibfnamefont {C.~D.}\ \bibnamefont {Marciniak}}, \bibinfo {author} {\bibfnamefont {S.}~\bibnamefont {Heu\ss{}en}}, \bibinfo {author} {\bibfnamefont {R.}~\bibnamefont {Blatt}}, \bibinfo {author} {\bibfnamefont {P.}~\bibnamefont {Schindler}}, \bibinfo {author} {\bibfnamefont {M.}~\bibnamefont {Rispler}}, \bibinfo {author} {\bibfnamefont {M.}~\bibnamefont {M\"uller}},\ and\ \bibinfo {author} {\bibfnamefont {T.}~\bibnamefont {Monz}},\ }\bibfield  {title} {\bibinfo {title} {Demonstration of fault-tolerant steane quantum error correction},\ }\href {https://doi.org/10.1103/PRXQuantum.5.030326} {\bibfield  {journal} {\bibinfo  {journal} {PRX Quantum}\ }\textbf {\bibinfo {volume} {5}},\ \bibinfo {pages} {030326} (\bibinfo {year} {2024})}\BibitemShut {NoStop}%
\bibitem [{\citenamefont {Huang}\ \emph {et~al.}(2024)\citenamefont {Huang}, \citenamefont {Brown},\ and\ \citenamefont {Cetina}}]{HuangSciAdv2024}%
  \BibitemOpen
  \bibfield  {author} {\bibinfo {author} {\bibfnamefont {S.}~\bibnamefont {Huang}}, \bibinfo {author} {\bibfnamefont {K.~R.}\ \bibnamefont {Brown}},\ and\ \bibinfo {author} {\bibfnamefont {M.}~\bibnamefont {Cetina}},\ }\bibfield  {title} {\bibinfo {title} {Comparing shor and steane error correction using the bacon-shor code},\ }\href {https://doi.org/10.1126/sciadv.adp2008} {\bibfield  {journal} {\bibinfo  {journal} {Science Advances}\ }\textbf {\bibinfo {volume} {10}},\ \bibinfo {pages} {eadp2008} (\bibinfo {year} {2024})},\ \Eprint {https://arxiv.org/abs/https://www.science.org/doi/pdf/10.1126/sciadv.adp2008} {https://www.science.org/doi/pdf/10.1126/sciadv.adp2008} \BibitemShut {NoStop}%
\bibitem [{\citenamefont {Zhang}\ \emph {et~al.}(2024)\citenamefont {Zhang}, \citenamefont {Wu},\ and\ \citenamefont {Guo}}]{ZhangPRR2024}%
  \BibitemOpen
  \bibfield  {author} {\bibinfo {author} {\bibfnamefont {J.}~\bibnamefont {Zhang}}, \bibinfo {author} {\bibfnamefont {Y.-C.}\ \bibnamefont {Wu}},\ and\ \bibinfo {author} {\bibfnamefont {G.-P.}\ \bibnamefont {Guo}},\ }\bibfield  {title} {\bibinfo {title} {Facilitating practical fault-tolerant quantum computing based on color codes},\ }\href {https://doi.org/10.1103/PhysRevResearch.6.033086} {\bibfield  {journal} {\bibinfo  {journal} {Phys. Rev. Res.}\ }\textbf {\bibinfo {volume} {6}},\ \bibinfo {pages} {033086} (\bibinfo {year} {2024})}\BibitemShut {NoStop}%
\bibitem [{\citenamefont {Lee}\ \emph {et~al.}(2025{\natexlab{a}})\citenamefont {Lee}, \citenamefont {Thomsen}, \citenamefont {Fazio}, \citenamefont {Brown},\ and\ \citenamefont {Bartlett}}]{Bartlett2025arXiv}%
  \BibitemOpen
  \bibfield  {author} {\bibinfo {author} {\bibfnamefont {S.-H.}\ \bibnamefont {Lee}}, \bibinfo {author} {\bibfnamefont {F.}~\bibnamefont {Thomsen}}, \bibinfo {author} {\bibfnamefont {N.}~\bibnamefont {Fazio}}, \bibinfo {author} {\bibfnamefont {B.~J.}\ \bibnamefont {Brown}},\ and\ \bibinfo {author} {\bibfnamefont {S.~D.}\ \bibnamefont {Bartlett}},\ }\href {https://arxiv.org/abs/2409.07707} {\bibinfo {title} {Low-overhead magic state distillation with color codes}} (\bibinfo {year} {2025}{\natexlab{a}}),\ \Eprint {https://arxiv.org/abs/2409.07707} {arXiv:2409.07707 [quant-ph]} \BibitemShut {NoStop}%
\bibitem [{\citenamefont {Delfosse}(2014)}]{DelfossePRA2014}%
  \BibitemOpen
  \bibfield  {author} {\bibinfo {author} {\bibfnamefont {N.}~\bibnamefont {Delfosse}},\ }\bibfield  {title} {\bibinfo {title} {Decoding color codes by projection onto surface codes},\ }\href {https://doi.org/10.1103/PhysRevA.89.012317} {\bibfield  {journal} {\bibinfo  {journal} {Phys. Rev. A}\ }\textbf {\bibinfo {volume} {89}},\ \bibinfo {pages} {012317} (\bibinfo {year} {2014})}\BibitemShut {NoStop}%
\bibitem [{\citenamefont {Sahay}\ and\ \citenamefont {Brown}(2022)}]{KaavyaPRXQ2022}%
  \BibitemOpen
  \bibfield  {author} {\bibinfo {author} {\bibfnamefont {K.}~\bibnamefont {Sahay}}\ and\ \bibinfo {author} {\bibfnamefont {B.~J.}\ \bibnamefont {Brown}},\ }\bibfield  {title} {\bibinfo {title} {Decoder for the triangular color code by matching on a m\"obius strip},\ }\href {https://doi.org/10.1103/PRXQuantum.3.010310} {\bibfield  {journal} {\bibinfo  {journal} {PRX Quantum}\ }\textbf {\bibinfo {volume} {3}},\ \bibinfo {pages} {010310} (\bibinfo {year} {2022})}\BibitemShut {NoStop}%
\bibitem [{\citenamefont {Gidney}\ and\ \citenamefont {Jones}(2023)}]{gidney2023arxiv}%
  \BibitemOpen
  \bibfield  {author} {\bibinfo {author} {\bibfnamefont {C.}~\bibnamefont {Gidney}}\ and\ \bibinfo {author} {\bibfnamefont {C.}~\bibnamefont {Jones}},\ }\href {https://arxiv.org/abs/2312.08813} {\bibinfo {title} {New circuits and an open source decoder for the color code}} (\bibinfo {year} {2023}),\ \Eprint {https://arxiv.org/abs/2312.08813} {arXiv:2312.08813 [quant-ph]} \BibitemShut {NoStop}%
\bibitem [{\citenamefont {Lee}\ \emph {et~al.}(2025{\natexlab{b}})\citenamefont {Lee}, \citenamefont {Li},\ and\ \citenamefont {Bartlett}}]{LeeQuantum2025}%
  \BibitemOpen
  \bibfield  {author} {\bibinfo {author} {\bibfnamefont {S.-H.}\ \bibnamefont {Lee}}, \bibinfo {author} {\bibfnamefont {A.}~\bibnamefont {Li}},\ and\ \bibinfo {author} {\bibfnamefont {S.~D.}\ \bibnamefont {Bartlett}},\ }\bibfield  {title} {\bibinfo {title} {Color code decoder with improved scaling for correcting circuit-level noise},\ }\href {https://doi.org/10.22331/q-2025-01-27-1609} {\bibfield  {journal} {\bibinfo  {journal} {{Quantum}}\ }\textbf {\bibinfo {volume} {9}},\ \bibinfo {pages} {1609} (\bibinfo {year} {2025}{\natexlab{b}})}\BibitemShut {NoStop}%
\bibitem [{\citenamefont {Lee}\ \emph {et~al.}(2025{\natexlab{c}})\citenamefont {Lee}, \citenamefont {Li},\ and\ \citenamefont {Bartlett}}]{Lee2025Github}%
  \BibitemOpen
  \bibfield  {author} {\bibinfo {author} {\bibfnamefont {S.-H.}\ \bibnamefont {Lee}}, \bibinfo {author} {\bibfnamefont {A.}~\bibnamefont {Li}},\ and\ \bibinfo {author} {\bibfnamefont {S.~D.}\ \bibnamefont {Bartlett}},\ }\bibfield  {title} {\bibinfo {title} {Color code decoder with improved scaling for correcting circuit-level noise},\ }\href {https://doi.org/10.22331/q-2025-01-27-1609} {\bibfield  {journal} {\bibinfo  {journal} {{Quantum}}\ }\textbf {\bibinfo {volume} {9}},\ \bibinfo {pages} {1609} (\bibinfo {year} {2025}{\natexlab{c}})}\BibitemShut {NoStop}%
\bibitem [{\citenamefont {Pelaez}\ \emph {et~al.}(2024)\citenamefont {Pelaez}, \citenamefont {Omole}, \citenamefont {Gokhale}, \citenamefont {Rines}, \citenamefont {Smith}, \citenamefont {Perlin},\ and\ \citenamefont {Hashim}}]{PelaezArXiv2024}%
  \BibitemOpen
  \bibfield  {author} {\bibinfo {author} {\bibfnamefont {E.}~\bibnamefont {Pelaez}}, \bibinfo {author} {\bibfnamefont {V.}~\bibnamefont {Omole}}, \bibinfo {author} {\bibfnamefont {P.}~\bibnamefont {Gokhale}}, \bibinfo {author} {\bibfnamefont {R.}~\bibnamefont {Rines}}, \bibinfo {author} {\bibfnamefont {K.~N.}\ \bibnamefont {Smith}}, \bibinfo {author} {\bibfnamefont {M.~A.}\ \bibnamefont {Perlin}},\ and\ \bibinfo {author} {\bibfnamefont {A.}~\bibnamefont {Hashim}},\ }\href {https://arxiv.org/abs/2403.12857} {\bibinfo {title} {Average circuit eigenvalue sampling on nisq devices}} (\bibinfo {year} {2024}),\ \Eprint {https://arxiv.org/abs/2403.12857} {arXiv:2403.12857 [quant-ph]} \BibitemShut {NoStop}%
\end{thebibliography}
\end{document}